\documentclass[onecolumn,tightenlines,nofootinbib,preprintnumbers,amsmath,amssymb]{revtex4}
\usepackage[utf8]{inputenc}
\usepackage[usenames,dvipsnames]{color}
\usepackage{caption2}
\usepackage{dcolumn}
\usepackage{graphicx}% Include figure files
\usepackage{subfigure}
\usepackage{epstopdf}
\usepackage{bm}% bold math
\usepackage{lscape}

\usepackage{verbatim}

\usepackage{setspace}
\begin{document}
\begin{spacing}{1.5}
%\nofiles

\title{Direct $CP$ violation of three bodies decay process from the resonance effect}

% Force line breaks with \\

\author{Gang L\"{u}$^{1}$\footnote{Email: ganglv66@sina.com},
Yan-Lin Zhao \footnote{Email: zyl163mail@163.com},
Liang-Chen Liu $^{1}$\footnote{Email: llc163mail@163.com},
Xin-Heng Guo $^{2}$\footnote{Email: xhguo@bnu.edu.cn}}

\affiliation{\small $^{1}$College of Science, Henan University of Technology, Zhengzhou 450001, China\\
 $^{2}$ College of Nuclear Science and Technology, Beijing Normal University, Beijing 100875, China \\
}

%\date{\today}

\begin{abstract}
The physical state of
$\rho$-$\omega$-$\phi$ mesons can be mixed by the unitary matrix.
The decay processes of $\omega \rightarrow \pi^{+}\pi^{-}$ and $\phi \rightarrow \pi^{+}\pi^{-}$
are from the isospin symmetry breaking.
The $\rho-\omega$, $\rho-\phi$ and $\omega-\phi$ interferences lead to resonance contribution
to produce the strong phases.
The $CP$ asymmetry is considered from above isospin symmetry breaking due to
the new strong phase for the first order.
It has been found the $CP$ asymmetry can be enhanced greatly for the decay process of $B^{0}\rightarrow\pi^+\pi^{-}\eta^{(')}$ when the invariant masses of the $\pi^+\pi^{-}$ pairs are in the area around the $\omega$ resonance range and the $\phi$ resonance range
in perturbative QCD. We also discuss the possibility to search the predicted $CP$ violation at the LHC.
\end{abstract}

\maketitle

\section{\label{intro}Introduction}
$CP$ violation reflects the asymmetry between matter and antimatter. Since the discovery of $CP$ violation in the decay process of K meson, theories and experiments have been exploring and searching for the source of $CP$ violation. In the standard model (SM), the weak complex phase of the Cabibbo-Kobayashi Maskawa (CKM) matrix is the main source of $CP$ violation in the process of particle weak decay \cite{cab}. The strong phase does not change under the $CP$ conjugate transformation. Because of the large mass of $B$-meson containing b-quark, the approximate result of perturbation calculation is good, which becomes an important field to search for $CP$ violation. In particular, in the two-body decay process of $B$-meson, the ratio of the penguin amplitude to the tree amplitude contributes the weak phase angle needed for $CP$ violation. Combining with the results of factorization, a relatively reliable prediction is given theoretically and measured experimentally. For example, the $CP$ violations in the decay processes of $B^{0}\rightarrow K^{+}\pi^{-} $ and $B^{0}\rightarrow \pi^{+}\pi^{-} $ are recently given \cite{Lees2013,DUH2013,Dalseno2013}. Compared with two-body decay process, three-body or multi-body decay contains more dynamic effects and the phase space distribution. With the measurement of LHCb Collaboration for $CP$ violation \cite{Aaijprl2013, Aaijprl2014}, multi-body decay process has become a research hotspot in recent years \cite{Raiprdl2014, Hsu2017, cheng2016, Klein2017, Lu2018, RA2019, RA2020, Aij2020}.

Experimentally, through model-independent analysis, a large $CP$ violation in the localized phase space region has been observed in the $B$-meson three-body noncharmed decay process \cite{Aaijprl2013,Aaijprl2014,Raiprdl2014,Hsu2017}, and there is no precise model to give the effect of resonance. In the recent literature \cite{RA2019}, the $CP$ violations observed in the decay process of $B^{\pm} \rightarrow K^{+}K^{-}\pi^{\pm}$ are given by the contributions of resonance, non-resonance and the $\pi\pi\rightarrow K\bar{K}$ re-scattering of final state particles. It is suggested that the re-scattering of final-state particles should play an equally important role in the other three-body non-charm-decay processes of $B$-meson. Since the isospin is conserved in the decay process of $\rho^{0}\rightarrow \pi^{+}\pi^{-}$ and the decay rate is 100$\%$, the large contribution of S-wave amplitude has been observed \cite{17Bau2005,18Bau2009}. At the same time, $CP$ violation was found in the low invariant mass region of S wave, which established that $CP$ violation was related to the interference of S wave and P wave amplitudes \cite{RA2020}. Some progress has also been made in the measurement of $CP$ violating phase angle and for the analysis of amplitudes in the $B$ meson four-body decay process. In the invariant mass regions of $K^{\pm}\pi^{\mp}$, the LHCb Collaboration analyzed the time-dependent amplitudes and phase angles of $CP$ violation by resonance of $K_{0}^{*}(800)^{0}$ and $K_{0}^{*}(1430)^{0}$,  $K_{0}^{*}(892)^{0}$ and $K_{2}^{*}(1430)^{0}$ \cite{19LHC2018}. Considering the resonance of $\rho$, $\omega$, $f_{0}(500)$,  $f_{0}(980)$ in the invariant mass regions of $\pi^{+}\pi^{-}$, and the main contribution of $K^{*}(892)^{0}$ decay  in the invariant mass regions of $K^{+}\pi^{-}$, one analyses the decay amplitude of $B^{0}\rightarrow(\pi^{+}\pi^{-})(K^{+}\pi^{-})$\cite{20LHC2019}.

Theoretically, three-body or multi-body decay is a relatively complex calculation process, and more studies have been done recently \cite{21Cheng2005, 22Bediaga2014, 23Bhattacharya2013, 24Xu2014, 25Bhattacharya2014, 26Cheng2016, 27Ma2018, 28Wang2019, 29Wang2016, 30Ye2017, 31Sheng2019}.
 Under the perturbation QCD framework, the final state interaction is described by the two-particle distribution amplitude in the resonance region. The three-body decay process is treated as a quasi-two-body decay process in the form of intermediate resonant states \cite{27Ma2018, 28Wang2019}.
Recently, the narrow width approximation is applied to extract the branching fraction
of the quasi-two-body decay processes by an intermediate resonant state.
The correction will be considered when the resonance has a sufficiently large width.
Since the widths $\omega(782)$ and $\phi(1020)$  are relatively small,
one can neglect the effects safely as quasi-two-body decay processes.
It has been show that the correction is generally less than $10\%$ for vector resonances.
It is known that the decay width of $\rho(770)$ is large.
The correction factor is at $7\%$ level for the decay process of
$B^{-}\rightarrow \rho(770)\pi^{-} \rightarrow \pi^{+}\pi^{-}\pi^{-}$ in the frame of QCD factorization
Under different factorization frameworks, the numerical results
may vary greatly within the error range. 
Particularly, we notice that the parameter $\eta_{R}$ is introduced to identify the degree of  approximation for $\Gamma(B \rightarrow RP_{3})B(R\rightarrow P_{1}P_{2})
=\eta_{R}\Gamma(B\rightarrow RP_{3}\rightarrow P_{1}P_{2}P_{3})$
in narrow width approximation \cite{haiyang2021prd, haiyang2021plb}. One can find that
the $\eta_{R}$ can be divided out for the calculation of $CP$ violation.
Based on the above considerations, we focus on vector meson resonance and ignore the effect of this correction in this work.

Considering the influence of isospin symmetry breaking, $CP$ violations for the decay processes of the three-body or four-body decay from $B$ meson have been studied via $\rho-\omega$ mixing \cite{29Wang2016, 30Ye2017, 31Sheng2019}. The mechnism of $\rho-\omega$ mixing produces strong phase to change the
$CP$ violation. Hence, the $\rho-\omega$, $\rho-\phi$ and $\omega-\phi$ interferences may lead to resonance contribution to produce the new strong phases. The $CP$ violation is considered from above isospin symmetry breaking due to the new strong phase for the first order. We will focus on
the $CP$ violations from the  $\rho-\omega-\phi$ interferences.

 The paper is organized as follows. In the second part, we give our theoretical derivation process
 for the resonance effects in detail. In the third part, we give the values of $CP$ violation in the decay process of individual decay process under our theoretical framework.  Summary and discussion are presented in forth part.

\section{\label{sec:cpv1}$CP$ violation from the resonance effects}
\subsection{\label{subsec:form}Formalism}

According to the vector mesons dominance (VMD)\cite{Kroll1967}, $e^{+}e^{-}$ annihilate into photons, which are polarized in vacuum to form vector particles $\rho^{0}(770)$, $\omega(782)$, $\phi(1020)$ and then decay into $\pi^{+}\pi^{-}$ pairs. The mixed amplitude parameters of the corresponding two or three particles can be obtained by the electromagnetic form factor of $\pi$ meson, and the values are given by combining with the experimental results \cite{connell1997}.
The intermediate state particle is a non-physical state, which is transformed into a physical field through an isospin field and connected by the unitary matrix R.
The mixed amplitude parameters can be expressed as $\Pi_{\rho\omega}$, $\Pi_{\rho\phi}$ and  $\Pi_{\omega\phi}$, and the contribution of higher order terms are ignored.
The momentum is transmitted by the vector meson by VMD model. These amplitudes should be related to the square of momentum. The transformation amplitudes are dependent on s associated with the square of momentum. The unitary matrix $R(s)$ relates the isopin field $\rho_{I}^{0}$, $\omega_{I}$, $\phi_{I}$  to the physical field $\rho^{0}$, $\omega$, $\phi$ by the relation:
\begin{equation}
\left (
\begin{array}{lllll}
\rho^0\\[0.5cm]
\omega\\[0.5cm]
\phi
\end{array}
\right )
=
R(s)
\left (
\begin{array}{lll}
\rho^0_I\\[0.5cm]
\omega_I\\[0.5cm]
\phi_I
\end{array}
\right ),
\label{L1}
\end{equation}
\noindent
 where
\begin{equation}
R  =
\left (
\begin{array}{lll}
<\rho_{I}|\rho> & \hspace{2.cm} <\omega_{I}|\rho>  &\hspace{2.cm}<\phi_{I}|\rho>\\[0.5cm]
<\rho_{I}|\omega> &  \hspace{2.cm}<\omega_{I}|\omega>&\hspace{2.cm}<\phi_{I}|\omega>\\[0.5cm]
<\rho_{I}|\phi>&\hspace{2.cm} <\omega_{I}|\phi> & \hspace{2.cm} <\phi_{I}|\phi>
\end{array}
\right )\\
\label{L2}
\end{equation}
\begin{equation}
=
\left (
\begin{array}{lll}
 ~~~~1 &\hspace{2.cm} -F_{\rho\omega}(s) &
\hspace{2.cm}  -F_{\rho\phi}(s)\\[0.5cm]
\displaystyle  F_{\rho\omega}(s) &  \hspace{2.cm}~~~ 1 &
\hspace{2.cm} \displaystyle - F_{\omega\phi}(s)\\[0.5cm]
\displaystyle   F_{\rho\phi}(s)
& \hspace{2.cm}  \displaystyle   F_{\omega\phi}(s)& \hspace{2.5cm} 1
\end{array}
\right ),
\label{L3}
\end{equation}
which $F_{\rho\omega}(s)$, $F_{\rho\phi}(s)$, $F_{\omega\phi}(s)$ is order $\mathcal{O}(\lambda)$,
$(\lambda\ll 1)$. The transformation of two representations is related to each other through unitary matrices R.
Based on isospin $\rho_{I}^{0}$, $\omega_{I}$, $\phi_{I}$ field, we can construct the isospin basis vector $|I,I_{3}>$. Thus, the physical particle state can be represented as a linear combination of the above basis vectors. We use M and N to represent the physical state and the isospin basis vector of the particle, respectively. According to the orthogonal normalization relation, we can get:
\begin{equation}
\sum\limits_{M}|M><M|=\sum\limits_{M_{I}}|M_{I}><M_{I}|=I,
\end{equation}
and
\begin{equation}
<M|N>=<M_{I}|N_{I}>=\delta_{MN}.
\end{equation}
One can write $|M>=\sum\limits_{N_{I}}|N_{I}><N_{I}|M>$ due to the
the transformation of two representations. $W$ can be defined as the mass squared operator.
The propagator can be defined as $D(s)=\frac{1}{s-W(s)}$ in the physical representation, which can
be written as $D(s)=\sum\limits_{M,N}|M><M|\frac{1}{s-W(s)}|N><N|$.
According to the diagonalization for the physical states, we obtain $<M|W|N>=\delta_{MN}Z_{M}$ to
generate $D=\sum\limits_{M}\frac{|M><M|}{s-Z_{M}}$ \cite{connell1997}.
From the translation of the two representations, the physical states can be written as
\begin{equation}
\rho^{0}=\rho^{0}_{I}-F_{\rho\omega}(s)\omega_{I}-F_{\rho\phi}(s)\phi_{I},
\end{equation}
\begin{equation}
\omega=F_{\rho\omega}(s)\rho^{0}_{I}+\omega_{I}-F_{\omega\phi}(s)\phi_{I},
\end{equation}
\begin{equation}
\phi=F_{\rho\Phi}(s)\rho^{0}_{I}+F_{\omega\phi}(s)\omega_{I}+\phi_{I}.
\end{equation}
We define
\begin{equation}
W_{I} =
\left (
\begin{array}{lll}
<\rho_{I}|W|\rho_{I}> & \hspace{2.cm} <\rho_{I}|W|\omega_{I}>  &\hspace{2.cm}<\rho_{I}|W|\phi_{I}>\\[0.5cm]
<\omega_{I}|W|\rho_{I}> &  \hspace{2.cm}<\omega_{I}|W|\omega_{I}>&\hspace{2.cm}<\omega_{I}|W|\phi_{I}>\\[0.5cm]
<\phi_{I}|W|\rho_{I}>&\hspace{2.cm} <\phi_{I}|W|\omega_{I}> & \hspace{2.cm} <\phi_{I}|W|\phi_{I}>
\end{array}
\right ).\\
\label{L2}
\end{equation}
Ignoring the contribution of higher order terms, we can diagonalize the equation $W_{I}$ by the matrix R in the physical representation.
\begin{equation}
W=RW_{I}R^{-1}=\left (
\begin{array}{lll}
Z_{\rho} & \hspace{2.cm}0  &\hspace{2.cm} 0  \\[0.5cm]
0 &  \hspace{2.cm}Z_{\omega}&\hspace{2.cm} 0\\[0.5cm]
0&\hspace{2.cm} 0 & \hspace{2.cm} Z_{\phi}
\end{array}
\right ).\\
\label{L3}
\end{equation}
From the Eqs.$(\ref{L2})$ and $(\ref{L3})$, we can neglect the high order terms of $F^{2}$, $F <\rho_{I}|W|\omega_{I}>$ and
 $F <\rho_{I}|W|\phi_{I}>$ for simplification. We can obtain the symmetry relationship
 of  $F <\rho_{I}|W|\omega_{I}>=F <\omega_{I}|W|\rho_{I}>$ and
 $F <\rho_{I}|W|\phi_{I}>=F <\phi_{I}|W|\rho_{I}>$ :
\begin{equation}
F_{\rho\omega}=\frac{<\rho_{I}|W|\omega_{I}>}{Z_{\omega}-Z_{\rho}},
\end{equation}
\begin{equation}
F_{\rho\phi}=\frac{<\rho_{I}|W|\phi_{I}>}{Z_{\phi}-Z_{\rho}},
\end{equation}
\begin{equation}
F_{\omega\phi}=\frac{<\omega_{I}|W|\phi_{I}>}{Z_{\phi}-Z_{\omega}}.
\end{equation}
The square of the complex mass can be written as \cite{connell1997, Harte1964}
\begin{equation}
Z_{\rho(\omega,\phi)}=(m_{\rho(\omega,\phi)}-i\Gamma_{\rho(\omega,\phi)}/2)^{2}\backsimeq m_{\rho(\omega,\phi)}^{2}-im_{\rho(\omega,\phi)}\Gamma_{\rho(\omega,\phi)},
\end{equation}
where $\Gamma_{\rho}$, $\Gamma_{\omega}$ and $\Gamma_{\phi}$ are the decay width of the mesons $\rho^0$, $\omega$, $\phi$, respectively.

Hence,
\begin{equation}
F_{\rho\omega}=\frac{<\rho_{I}|W|\omega_{I}>}{m_{\omega}^{2}-m_{\rho}^{2}-i(m_{\omega}\Gamma_{\omega}-m_{\rho}\Gamma_{\rho})},
\end{equation}
\begin{equation}
F_{\rho\phi}=\frac{<\rho_{I}|W|\phi_{I}>}{m_{\phi}^{2}-m_{\rho}^{2}-i(m_{\phi}\Gamma_{\phi}-m_{\rho}\Gamma_{\rho})},
\end{equation}
\begin{equation}
F_{\omega\phi}=\frac{<\omega_{I}|W|\phi_{I}>}{m_{\phi}^{2}-m_{\omega}^{2}-i(m_{\phi}\Gamma_{\phi}-m_{\omega}\Gamma_{\omega})}.
\end{equation}

In the physical representation, the propagator of intermediate state particle from vector meson can be expressed as
\begin{equation}
D_{V_{1}V_{2}}^{\mu\nu}(q^{2})=i\int d^{4}x e^{iqx}<0|T(V_{1}^{\mu}(x)(V_{2}^{\nu}(0))|0>.
\end{equation}
$D_{V_{1}V_{2}}$ and
$D_{V_{1}V_{2}}^{I}$ refer to the propagator $D_{V_{1}V_{2}}=<0|TV_{1}V_{2}|0>$ and $D_{V_{1}V_{2}}^{I}=<0|TV_{1}^{I}V_{2}^{I}|0>$  in the representations of physics and isospin, respectively.
We can obtain
\begin{eqnarray}
D_{\rho\omega}&=&<0|T\rho\omega|0>=<0|T(\rho_{I}-F_{\rho\omega}\omega_{I}-F_{\rho\phi}\phi_{I})(F_{\rho\omega}\rho_{I}+\omega_{I}-F_{\omega\phi}\phi_{I})|0> \\ \nonumber
&=& F_{\rho\omega}\frac{1}{s_{\rho}}+\frac{1}{s_{\rho}}\Pi_{\rho\omega}\frac{1}{s_{\omega}}-F_{\omega\phi}\frac{1}{s_{\rho}}\Pi_{\rho\phi}\frac{1}{s_{\phi}}
-F_{\rho\omega}\frac{1}{s_{\omega}}-F_{\rho\phi}\frac{1}{s_{\phi}}\Pi_{\phi\omega}\frac{1}{s_{\omega}}+\mathcal{O}(\varepsilon^{2}),
\end{eqnarray}
\begin{eqnarray}
D_{\rho\phi}&=&<0|T\rho\phi|0>=<0|T(\rho_{I}-F_{\rho\omega}\omega_{I}-F_{\rho\phi}\phi_{I})(F_{\rho\phi}\rho_{I}+F_{\omega\phi}\omega_{I}+\phi_{I})|0> \\ \nonumber
&=& F_{\rho\phi}\frac{1}{s_{\rho}}+F_{\omega\phi}\frac{1}{s_{\rho}}\Pi_{\rho\omega}\frac{1}{s_{\omega}}+\frac{1}{s_{\rho}}\Pi_{\rho\phi}\frac{1}{s_{\phi}}
-F_{\rho\omega}\frac{1}{s_{\omega}}\Pi_{\omega\phi}\frac{1}{s_{\phi}}-F_{\rho\phi}\frac{1}{s_{\phi}}+\mathcal{O}(\varepsilon^{2}) , \\ \nonumber
\end{eqnarray}
\begin{eqnarray}
D_{\omega\phi}&=&<0|T\omega\phi|0>=<0|T(\omega_{I}+F_{\rho\omega}\rho_{I}-F_{\omega\phi}\phi_{I})(F_{\rho\phi}\rho_{I}+F_{\omega\phi}\omega_{I}+\phi_{I})|0> \\ \nonumber
&=&F_{\rho\omega}\frac{1}{s_{\rho}}\Pi_{\rho\omega}\frac{1}{s_{\phi}}+F_{\rho\phi}\frac{1}{s_{\omega}}\Pi_{\omega\rho}\frac{1}{s_{\rho}}+F_{\omega\phi}\frac{1}{s_{\omega}}
+\frac{1}{s_{\omega}}\Pi_{\omega\phi}\frac{1}{s_{\phi}}-F_{\omega\phi}\frac{1}{s_{\phi}}+\mathcal{O}(\varepsilon^{2}).   \\ \nonumber
\end{eqnarray}
In the same way, $D_{\omega\rho}=D_{\rho\omega}$, $D_{\rho\phi}=D_{\phi\rho}$ and $D_{\omega\phi}=D_{\phi\omega}$.

In the state of physics, there are not the $\rho-\omega-\phi$ mixing  so that $D_{\rho\omega}$,  $D_{\rho\phi}$ and $D_{\omega\phi}$ are equal to zero.
One can get
\begin{eqnarray}
\frac{1}{s_{\rho}}\Pi_{\rho\omega}\frac{1}{s_{\omega}}-F_{\omega\phi}\frac{1}{s_{\rho}}\Pi_{\rho\phi}\frac{1}{s_{\phi}}
-F_{\rho\phi}\frac{1}{s_{\phi}}\Pi_{\phi\omega}\frac{1}{s_{\omega}}=F_{\rho\omega}(\frac{1}{s_{\omega}}-\frac{1}{s_{\rho}}),
\end{eqnarray}
\begin{eqnarray}
F_{\omega\phi}\frac{1}{s_{\rho}}\Pi_{\rho\omega}\frac{1}{s_{\omega}}+\frac{1}{s_{\rho}}\Pi_{\rho\phi}\frac{1}{s_{\phi}}
-F_{\rho\omega}\frac{1}{s_{\omega}}\Pi_{\omega\phi}\frac{1}{s_{\phi}}=F_{\rho\phi}(\frac{1}{s_{\phi}}-\frac{1}{s_{\rho}}),
\end{eqnarray}
\begin{eqnarray}
F_{\rho\omega}\frac{1}{s_{\rho}}\Pi_{\rho\omega}\frac{1}{s_{\phi}}+F_{\rho\phi}\frac{1}{s_{\omega}}\Pi_{\omega\rho}\frac{1}{s_{\rho}}
+\frac{1}{s_{\omega}}\Pi_{\omega\phi}\frac{1}{s_{\phi}}=F_{\omega\phi}(\frac{1}{s_{\phi}}-\frac{1}{s_{\omega}}).
\end{eqnarray}

The parameters of $\Pi_{\rho\omega}$, $\Pi_{\omega\phi}$, $\Pi_{\rho\phi}$, $F_{\rho\omega}$, $F_{\rho\phi}$ and $F_{\omega\phi}$
are order of $\mathcal{O}(\lambda)$ ($\lambda\ll 1$).
Any two or three terms multiplied together are of higher order and can be ignored.
Hence, we can get
\begin{eqnarray}
F_{\rho\omega}=\frac{\Pi_{\rho\omega}}{s_{\rho}-s_{\omega}},
\end{eqnarray}
\begin{eqnarray}
F_{\rho\phi}=\frac{\Pi_{\rho\phi}}{s_{\rho}-s_{\phi}},
\end{eqnarray}
and
\begin{eqnarray}
F_{\omega\phi}=\frac{\Pi_{\omega\phi}}{s_{\omega}-s_{\phi}},
\end{eqnarray}
where we can define
\begin{eqnarray}
\widetilde{\Pi}_{\rho\omega}=\frac{s_{\rho}\Pi_{\rho\omega}}{s_{\rho}-s_{\omega}},
\end{eqnarray}
\begin{eqnarray}
\widetilde{\Pi}_{\rho\phi}=\frac{s_{\rho}\Pi_{\rho\phi}}{s_{\rho}-s_{\phi}}.
\end{eqnarray}
$s_{V}$, $m_{V}$, and $\Gamma_{V}$ ($V$= $\rho$, $\omega$ or $\phi$) refer to the inverse propagator, mass and decay rate of the vector meson $V$, respectively. We can write
\begin{eqnarray}
s_V=s-m_V^2+{\rm{i}}m_V\Gamma_V,
\end{eqnarray}
where the $\sqrt{s}$ denotes the invariant mass of the $\pi^+ \pi^-$ pairs \cite{AG}.

The $\rho-\omega$ mixing paraments were recently determined precisely by Wolfe and Maltnan \cite{Wolf2009,Wolf2011}
\begin{eqnarray}
\begin{split}
& \quad
\mathfrak{Re}{\Pi}_{\rho\omega}(m_{\rho}^2)&=&-4470\pm250_{model}\pm160_{data}\rm{MeV}^2,\\
& \quad
{\mathfrak{Im}}{\Pi}_{\rho\omega}(m_{\rho}^2)&=&-5800\pm2000_{model}\pm1100_ {data}\textrm{MeV}^2.
\end{split}
\end{eqnarray}

The $\rho-\phi$ mixing paraments have been given near the $\phi$ meson \cite{ACH2000}
\begin{eqnarray}
F_{\rho\phi}=(0.72\pm 0.18)\times 10^{-3}-i(0.87 \pm 0.32)\times 10^{-3}.
\end{eqnarray}
The mixing parameter depends on the momentum including both the resonant and non-resonant contribution which absorbs the direct decay processes $\omega\rightarrow\pi^+\pi^-$
and $\phi\rightarrow\pi^+\pi^-$ from the isospin symmetry breaking effects.
The mixing parameters $\widetilde{\Pi}_{\rho\omega}(s)$ and $\widetilde{\Pi}_{\rho\phi}(s)$
are the momentum dependence for $\rho-\omega$ mixing and $\rho-\phi$ mixing, respectively.
We expect to search for the contribution of this mixing mechanism in the resonance region
of $\omega$ and $\phi$ mass where two pions are also produced by isospin symmetry breaking.
One can express $\widetilde{\Pi}_{\rho\omega}(s)={\mathfrak{Re}}\widetilde{\Pi}_{\rho\omega}(m_{\omega}^2)+{\mathfrak{Im}}\widetilde{\Pi}_{\rho\omega}(m_{\omega}^2)$
and $\widetilde{\Pi}_{\rho\phi}(s)={\mathfrak{Re}}\widetilde{\Pi}_{\rho\phi}(m_{\phi}^2)+{\mathfrak{Im}}\widetilde{\Pi}_{\rho\phi}(m_{\phi}^2)$
and update the values:
\begin{eqnarray}
\begin{split}
\mathfrak{Re}\widetilde{\Pi}_{\rho\omega}(m_{\omega}^2)&=&-4760\pm440
\rm{MeV}^2,\\
{\mathfrak{Im}}\widetilde{\Pi}_{\rho\omega}(m_{\omega}^2)&=&-6180\pm3300
\textrm{MeV}^2,
\end{split}
\end{eqnarray}
and
\begin{eqnarray}
\begin{split}
\mathfrak{Re}\widetilde{\Pi}_{\rho\phi}(m_{\phi}^2)&=&796\pm312
\rm{MeV}^2,\\
{\mathfrak{Im}}\widetilde{\Pi}_{\rho\phi}(m_{\phi}^2)&=&-101\pm67
\textrm{MeV}^2.
\end{split}
\end{eqnarray}

\subsection{\label{sec:cpv1}$CP$ violation in $B^{0}\rightarrow \rho^0(\omega,\phi)\eta^{(')}\rightarrow \pi^+\pi^{-}\eta^{(')}$}
%\subsection{\label{subsec:form}Formalism}
The experiments of $e^{+}e^{-} \rightarrow hadrons $
are measured for the cross section to determine the parameters of vector mesons
in the energy range of  $\rho^0$, $\omega$ and $\phi$ from
the reactions.
The processes of  $\omega \rightarrow \pi^{+}\pi^{-}$ and
$\phi \rightarrow \pi^{+}\pi^{-}$  from the isospin symmetry breaking
can provide dynamics information about
the interference of $\rho^0$, $\omega$ and $\phi$ mesons.
$CP$ violation depends on the CKM matrix elements associated with
weak phase and strong phase.
The effect of the isospin symmetry breaking
can provide the strong phase to change the $CP$ violation from
intermediate vector meson mixing.
We take the $B^{0}\rightarrow \rho^0(\omega,\phi)\eta^{(')}\rightarrow \pi^+\pi^{-}\eta^{(')}$
decay channel as an example to study $CP$ violation.

The decay amplitude $A$($\bar{A}$) for the process of
$B^{0}\rightarrow\pi^+\pi^{-}\eta^{(')}$
can be expressed as:
\begin{eqnarray}
A=\big<\pi^+\pi^{-}\eta^{(')}|H^T|B^{0}\big>+\big<\pi^+\pi^{-}\eta^{(')}|H^P|B^{0}\big>,
\label{A}
\end{eqnarray}
where $\big<\pi^+\pi^{-}\eta^{(')}|H^T|B^{0}\big>$ and $\big<\pi^+\pi^{-}\eta^{(')}|H^P|B^{0}\big>$
refer to the contribution from the tree level and
penguin level due to the operators of Hamiltonian, respectively.
The ratio of the penguin diagram contribute to the tree diagram contribution produces the phase angle, which affects the $CP$ violation in the decay process. The formalism of amplitude can be expressed as follows:
\begin{eqnarray}
A=\big<\pi^+\pi^{-}\eta^{(')}|H^T|B^{0}\big>[1+re^{i(\delta+\phi)}].
\label{A'}
\end{eqnarray}
The weak phase $\phi$ is from the CKM matrix.
The strong phase $\delta$ and parameter $r$
are dependent on the interference of the two level contribution and
other mechanism. We can define
\begin{eqnarray}
r\equiv\Bigg|\frac{\big<\pi^+\pi^{-}\eta^{(')}|H^P|B^{0}\big>}{\big<\pi^+\pi^{-}\eta^{(')}|H^T|B^{0}\big>}\Bigg|
\label{r}.
\end{eqnarray}
We can provide the decay amplitude from the isospin field.
Then, the physical decay amplitude
is obtained by the translation of the two representation
from the $B\rightarrow V_{I}$ and $V_{I}\rightarrow \pi^{+}\pi^{-}$ by the
unitary matrix $R$. One can find the propagators of intermediate vector mesons
become physical states from the diagonal matrix.
To the leading order approximation
of isospin violation, one can provide the following results:
\begin{eqnarray}
\big<\pi^+\pi^{-}\eta^{(')}|H^T|B^{0}\big>=\frac{g_{\rho}}{s_{\rho}s_{\omega}}\widetilde{\Pi}_{\rho\omega}t_{\omega}
+\frac{g_{\rho}}{s_{\rho}s_{\phi}}\widetilde{\Pi}_{\rho\phi}t_{\phi}+\frac{g_{\rho}}{s_{\rho}}t_{\rho},
\label{Htr}
\end{eqnarray}
\begin{eqnarray}
\big<\pi^+\pi^{-}\eta^{(')}|H^P|B^{0}\big>=\frac{g_{\rho}}{s_{\rho}s_{\omega}}\widetilde{\Pi}_{\rho\omega}p_{\omega}
+\frac{g_{\rho}}{s_{\rho}s_{\phi}}\widetilde{\Pi}_{\rho\phi}p_{\phi}+\frac{g_{\rho}}{s_{\rho}}p_{\rho},
\label{Hpe}
\end{eqnarray}
where $t_{\rho}(p_{\rho})$, $t_{\phi}(p_{\phi})$, and
$t_{\omega}(p_{\omega})$ are the tree (penguin) amplitudes
for $B^{0}\rightarrow\rho^{0}\eta^{(')}$, $B^{0}\rightarrow\phi\eta^{(')}$ and
$B^{0}\rightarrow\omega\eta^{(')}$, respectively.
The coupling constant $g_{\rho}$ is from decay process of $\rho^0\rightarrow\pi^+\pi^-$.
Then, we can obtain
\begin{eqnarray}
re^{i\delta}e^{i\phi}=\frac{\widetilde{\Pi}_{\rho\omega}p_{\omega}s_{\phi}+\widetilde{\Pi}_{\rho\phi}p_{\phi}s_{\omega}
+s_{\omega}s_{\phi}p_{\rho}}{\widetilde{\Pi}_{\rho\omega}t_{\omega}s_{\phi}+\widetilde{\Pi}_{\rho\phi}t_{\phi}s_{\omega}+s_{\omega}s_{\phi}t_{\rho}},
\label{rdtdirive}
\end{eqnarray}
Defining
\begin{eqnarray}
\frac{p_{\omega}}{t_{\rho}}\equiv r_{1}
e^{i(\delta_\lambda+\phi)},
\quad\frac{p_{\phi}}{t_{\rho}}\equiv r_{2}e^{i(\delta_\chi+\phi)},
\quad\frac{t_{\omega}}{t_{\rho}}\equiv \alpha e^{i\delta_\alpha},
\quad\frac{t_{\phi}}{t_{\rho}}\equiv \tau e^{i\delta_\tau},
\quad\frac{p_{\rho}}{p_{\omega}}\equiv \beta e^{i\delta_\beta}, \label{def}
\end{eqnarray}
with $\delta_\alpha$, $\delta_\beta$ and $\delta_q$ are strong phases. It is available from Eqs.(\ref{rdtdirive})(\ref{def}):
\begin{eqnarray}
re^{i\delta}=\frac{\widetilde{\Pi}_{\rho\omega} r_{1}
e^{i\delta_\lambda}s_{\phi}+\widetilde{\Pi}_{\rho\phi}r_{2}
e^{i\delta_\chi}s_{\omega}
+s_{\omega}s_{\phi}\beta e^{i\delta_\beta} r_{1}
e^{i\delta_\lambda}}{\widetilde{\Pi}_{\rho\omega}\alpha e^{i\delta_\alpha}s_{\phi}+\widetilde{\Pi}_{\rho\phi}\tau e^{i\delta_\tau}s_{\omega}+s_{\omega}s_{\phi}},
\end{eqnarray}

We need give sin$\phi$ and cos$\phi$ to obtain the $CP$ violation
The weak phase $\phi$ comes from the CKM matrix elements. In the Wolfenstein
parametrization \cite{wol}, one has
\begin{eqnarray}
\begin{split}
{\rm sin}\phi &=&\frac{\eta}{\sqrt{[\rho(1-\rho)-\eta^2]^2+\eta^2}},\\
{\rm cos}\phi &=&\frac{\rho(1-\rho)-\eta^2}{\sqrt{[\rho(1-\rho)-\eta^2]^2+\eta^2}}.
\label{3l1}
\vspace{2mm}
\end{split}
\end{eqnarray}

\subsection{\label{cal} The details of calculation}
In the Perturbative QCD method, on a rest frame of heavy $B$ meson,
the $B$ mesons decay into two light mesons by large momenta which move very fast.
The perturbative interaction plays a major role in the decay process over a short distance.
There is not enough time to exchange soft gluon among final state mesons.
Since the final state mesons move very fast, the hard gluon gives a lot of energy to the spectator quark of the $B$ meson which produces a fast moving final meson.
Non-perturbative contributions are included in the meson wave functions and form factors.
One can use perturbation theory to calculate the decay amplitudes
by introducing the Sudakov factor to eliminate endpoint divergence.

For simplification, we take the decay process
$B^{0}\rightarrow \rho^{0}(\omega,\phi)\eta \rightarrow \pi^{+}\pi^{-} \eta $
as an example to illustrate the mechanism in detail.
One need obtain the formalisms of $t_{\rho}$, $t_{\omega}$,  $t_{\phi}$
and $p_{\rho}$, $p_{\omega}$,  $p_{\phi}$ to calculate the $CP$ violations
which are from the tree level and penguin level contributions, respectively.
$C_{i}$ are the Wilson coefficients. One can find the formalisms of the functions F and M by Appendix.

Based on CKM matrix elements of $V_{ub}V^{*}_{ud}$ and
$V_{tb}V^{*}_{td}$, the decay amplitude of $B^{0}\rightarrow \rho^{0}\eta$ in
perturbation QCD approach can be written as
\begin{eqnarray}
\sqrt{6}M(B^{0}\to\rho^{0} \eta)=V_{ub}V^{*}_{ud}t_{\rho}-V_{tb}V^{*}_{td}p_{\rho},
\label{Brhoeta}
\end{eqnarray}
where $t_{\rho}$ and $p_{\rho}$ refer to the tree and penguin contributions respectively.
The formalisms can be obtained by the perturbative QCD method.

One can get
\begin{eqnarray}
\begin{split}
& \quad
t_{\rho}&=&F_{e}\left(C_{1}+\frac{1}{3}C_{2}\right) F_{1}(\theta_{p})
- F_{e\rho}\left(C_{1}+\frac{1}{3}C_{2} \right) f_{\eta}^{d} F_{1}(\theta_{p})\\
& \quad
 &-& M_{e\rho} C_{2}\cdot F_{1}(\theta_p)+\left (M_{a\rho}+M_{a}\right)
 C_{2} F_{1}(\theta_{p}) + M_e C_{2} F_{1}(\theta_{p}),
\end{split}
\end{eqnarray}
and
\begin{eqnarray}
\begin{split}
& \quad
-p_{\rho} &=& F_{e} \left [- \left (-\frac{1}{3}C_{3}
-C_{4}+\frac{3}{2}C_{7}+\frac{1}{2}C_{8}+\frac{5}{3}C_{9}
+C_{10}\right)\right] F_{1}(\theta_{p})   \\
& \quad
&&- F_{e\rho}\left [ -\left(\frac{1}{3}C_{3}+C_{4}
-\frac{1}{2}C_{7}-\frac{1}{6}C_{8}+\frac{1}{3}C_{9}-\frac{1}{3}
C_{10}\right)\right] f_{\eta}^d F_{1}(\theta_p)
\\
& \quad
&&
- F_{e\rho}\left(\frac{1}{2}C_7+\frac{1}{6}C_{8}
+\frac{1}{2}C_9+\frac{1}{6}C_{10}\right)
 f_{\eta}^s  F_{2}(\theta_p)   \\
 & \quad
&&
 + F_{e\rho}^{P} \left(\frac{1}{3}C_{5}+
C_{6}-\frac{1}{6}C_{7} -\frac{1}{2}C_{8} \right) \cdot F_{1}(\theta_p)
\\
& \quad
&& - M_{e\rho} \left [ - \left(C_3 + 2
C_4+2C_6+\frac{1}{2}C_8-\frac{1}{2}C_9
+\frac{1}{2}C_{10}\right)\right ] \cdot F_{1}(\theta_p) \\
& \quad
&& + M_{e\rho}\left(
C_4+C_6-\frac{1}{2}C_8-\frac{1}{2}C_{10}\right )
F_2(\theta_p) +\left (M_{a\rho}+M_a\right) \left
[-\left( -C_3+\frac{3}{2}C_8+\frac{1}{2}C_9
+\frac{3}{2}C_{10}\right)\right ] F_1(\theta_p) \\
& \quad
&&
- \left(M_e^{P}+2 M_a^{P}\right)
  \left (C_5-\frac{1}{2}C_7 \right )
 F_1(\theta_p)+ M_e \left[-\left(-C_3 -\frac{3}{2}C_8+\frac{1}{2}C_{9}
+\frac{3}{2}C_{10}\right)\right ] F_{1}(\theta_{p}),
\end{split}
\end{eqnarray}
where $F_{1}\left(\theta_{p}\right)=-\sin \theta_{p}+\cos \theta_{p} / \sqrt{2}$ and $F_{2}\left(\theta_{p}\right)=-\sin \theta_{p}-\sqrt{2} \cos \theta_{p}$, one can obtain $-20^{\circ} <\theta_{p}< -10^{\circ}$ as given in Ref. \cite{thetap}. The decay amplitudes for $B^{0}\to\rho^{0} \eta'$ can be obtained from Eqs.(\ref{Brhoeta}) by the following replacements
$f_{\eta}^{d}, f_{\eta}^{s} \rightarrow f_{\eta^{\prime}}^{d} f_{\eta^{\prime}}^{s} $,
$F_{1}\left(\theta_{p}\right) \rightarrow F_{1}^{\prime}\left(\theta_{p}\right)=\cos \theta_{p}+\frac{\sin \theta_{p}}{\sqrt{2}}$, and
$F_{2}\left(\theta_{p}\right) \rightarrow F_{2}^{\prime}\left(\theta_{p}\right)=\cos \theta_{p}-\sqrt{2} \sin \theta_{p}$, where the possible gluonic component of $\eta'$ meson has been neglected here.

The $t_{\omega}$ and $p_{\omega}$ can be extracted by the amplitudes of $B^0 \to \omega\eta $.
The decay amplitudes can be written as
\begin{eqnarray}
\sqrt{2}M(B^0\to \omega \eta)=V_{ub}V^{*}_{ud}t_{\omega}-V_{tb}V^{*}_{td}p_{\omega},
\label{Bomegaeta}
\end{eqnarray}
where
\begin{eqnarray}
\begin{split}
& \quad
t_\omega &=& F_{e}F_1(\phi) f_{\omega}\left( C_1 +
\frac{1}{3}C_2\right)
 + M_{e}F_1(\phi)C_2+F_{e\omega}
  \left( C_1 +\frac{1}{3}C_2\right)f_{\eta}^{d}
 \\ & \quad &
&+M_{e\omega}F_1(\phi)C_2 +
\left(M_{a}+M_{a\omega}\right) F_1(\phi)C_{2},
\end{split}
\end{eqnarray}
and
\begin{eqnarray}
\begin{split}
& \quad
-p_\omega &=&- F_{e}F_1(\phi) f_{\omega}
\left(\frac{7}{3}C_3+\frac{5}{3}C_4+2C_{5}+\frac{2}{3}C_{6}
+\frac{1}{2}C_7+\frac{1}{6}C_8+\frac{1}{3}C_9
 -\frac{1}{3} C_{10}\right)
 \\ & \quad
& & + M_{e}F_1(\phi)\left [-
\left(C_3+2C_4-2C_6-\frac{1}{2}C_8-\frac{1}{2}C_9
+\frac{1}{2}C_{10}\right)\right ]
  \\ & \quad && +F_{e\omega} \left[ -
\left(\frac{7}{3}C_3+\frac{5}{3}C_4-2C_{5}-\frac{2}{3}C_{6}
-\frac{1}{2}C_7-\frac{1}{6}C_8+\frac{1}{3}C_9
 -\frac{1}{3} C_{10}\right)f_{\eta}^{d}\right.\\ & \quad & &\left.
 -\left(C_3+\frac{1}{3}C_4-C_5-\frac{1}{3}C_6
 +\frac{1}{2}C_7+\frac{1}{6}C_8-\frac{1}{2}C_9-\frac{1}{6}C_{10}\right)f_{\eta}^{s}\right]
\\ & \quad
& & +F_{e\omega}^{P_2}\left[-\left(\frac{1}{3}C_5+C_6-
\frac{1}{6}C_7-\frac{1}{2}C_8\right)f_{\eta}^{d}\right] \\ & \quad &
&+M_{e\omega}F_1(\phi)\left [-
\left(C_3+2C_4+2C_6+\frac{1}{2}C_8-\frac{1}{2}C_9
+\frac{1}{2}C_{10}\right)\right ]
  \\ & \quad && + M_{e\omega}F_2(\phi)\left [-
\left(C_4+C_6-\frac{1}{2}C_8-\frac{1}{2}C_{10}\right)\right ]
 \\ & \quad &&+
\left(M_{a}+M_{a\omega}\right) F_1(\phi)\left[ -
\left(C_3+2C_4-\frac{1}{2}C_9 +\frac{1}{2}C_{10}\right)\right ] \\ & \quad
&&
 -\left(M_{e}^{P_1}\,+M_{a}^{P_1}\,+M_{a\omega}^{P_1}\,\right)F_1(\phi)
\,\left(C_{5}-\frac{1}{2}C_{7}\right)
\\ & \quad && -
 \left(M_{a}^{P_2}\,+
 M_{a\omega}^{P_2}\,\right) F_1(\phi)\,
\left(2C_6+\frac{1}{2}C_8\right).
\end{split}
\end{eqnarray}
By replacing the $f_{\omega} \to f_{\phi}$ and $\phi_{\omega}^{A,P,T} \to \phi_{\phi}^{A,P,T}$, the amplitude for $B^0 \to \phi\eta $ can be written as
\begin{eqnarray}
M(B^0\to \phi \eta)=V_{ub}V^{*}_{ud}t_{\phi}-V_{tb}V^{*}_{td}p_{\phi},
\label{Bphieta}
\end{eqnarray}
where
\begin{eqnarray}
t_{\phi}=0,
\end{eqnarray}
and
\begin{eqnarray}
\begin{split}
& \quad
p_{\phi} &=& F_{e}\;  \left(C_3+\frac{1}{3}C_4+C_{5}+\frac{1}{3}C_{6}
-\frac{1}{2}C_7-\frac{1}{6}C_8-\frac{1}{2}C_9
 -\frac{1}{6} C_{10}\right) F_1(\phi) \\ & \quad
& & + M_{e} \; \; \left(C_4-C_6+\frac{1}{2}C_8
-\frac{1}{2}C_{10}\right) \; F_1(\phi)
  +\left( M_{a}+ M_{a\phi}\right) \;
\left(C_4-\frac{1}{2}C_{10}\right) F_2(\phi) \\ & \quad &&
+\left(M_{a}^{P_2}+M_{a\phi}^{P_2}\right) \;  \,
\left(C_6-\frac{1}{2}C_8\right) F_2(\phi).
\end{split}
\end{eqnarray}
where $F_{1}(\phi)=\cos \phi / \sqrt{2}$ and $F_{2}(\phi)=-\sin \phi$, one can get $\phi = 39.3^{\circ}\pm1.0^{\circ}$ in Ref. \cite{bib20,bib21}. The complete decay amplitudes for $B^{0}\to\omega \eta'$ and $B^{0}\to\phi \eta'$ can be obtained from Eqs.(\ref{Bomegaeta}) and Eqs.(\ref{Bphieta}) by the following replacements
$f_{\eta}^{d}, f_{\eta}^{s} \rightarrow f_{\eta^{\prime}}^{d} f_{\eta^{\prime}}^{s} $,
$F_{1}\left(\phi\right) \rightarrow F_{1}^{\prime}\left(\phi\right)=\frac{\sin \phi}{\sqrt{2}}$, and
$F_{2}\left(\phi\right) \rightarrow F_{2}^{\prime}\left(\phi\right)=\cos \phi$.

\subsection{\label{input}INPUT PARAMETERS AND WAVE FUNCTIONS}
The CKM matrix, which elements are determined from experiments, can be expressed in terms of the Wolfenstein parameters $A$, $\rho$, $\lambda$ and $\eta$ \cite{wol}:
\begin{equation}
\left(
\begin{array}{ccc}
  1-\tfrac{1}{2}\lambda^2   & \lambda                  &A\lambda^3(\rho-\mathrm{i}\eta) \\
  -\lambda                 & 1-\tfrac{1}{2}\lambda^2   &A\lambda^2 \\
  A\lambda^3(1-\rho-\mathrm{i}\eta) & -A\lambda^2              &1\\
\end{array}
\right),\label{ckm}
\end{equation}
where $\mathcal{O} (\lambda^{4})$ corrections are neglected. The latest values for the parameters in the CKM matrix are \cite{PDG2020}:
\begin{eqnarray}
\begin{split}
&& \lambda=0.22650\pm0.00048,\quad A=0.790^{+0.017}_{-0.012}, \\
&& \bar{\rho}=0.141_{-0.017}^{+0.016},\hspace{0.8cm}\quad
\bar{\eta}=0.357\pm0.011,
\label{eq: rhobarvalue}
\end{split}
\end{eqnarray}
where
\begin{eqnarray}
 \bar{\rho}=\rho(1-\frac{\lambda^2}{2}),\quad
\bar{\eta}=\eta(1-\frac{\lambda^2}{2}).\label{eq: rho rhobarrelation}
\end{eqnarray}

From Eqs. (\ref{eq: rhobarvalue}) ( \ref{eq: rho rhobarrelation})
we have
\begin{eqnarray}
0.127<\rho<0.161,\quad  0.355<\eta<0.377.\label{eq: rho value}
\end{eqnarray}

The other parameters are given as following \cite{wol,PDG2020}:
\begin{eqnarray}
\begin{split}
&\quad
m_B=5.2792\text{GeV},\hspace{2cm} m_W=80.385\text{GeV},\\
&\quad
m_\rho=0.77526\text{GeV},\hspace{1.9cm} m_\phi=1.02\text{GeV},\\
&\quad
m_\omega=0.78265\text{GeV},\hspace{1.9cm}   C_{F}=4/3, \\
&\quad
f_\rho=0.216\text{GeV},\hspace{2.4cm} f^T_\rho=0.17\text{GeV},\\
&\quad
f_\omega=0.195\text{GeV},\hspace{2.3cm} f^T_\omega=0.14\text{GeV},\\
&\quad
f_\phi=0.237\text{GeV},\hspace{2.3cm} f^T_\phi=0.22\text{GeV},\\
&\quad
f_\pi=0.13\text{GeV},\hspace{2.5cm}\Gamma_\rho=0.15\text{GeV},\\
&\quad
\Gamma_\omega=8.49\times10^{-3}\text{GeV},\hspace{1.3cm}\Gamma_\phi =4.23\times10^{-3}\text{GeV},\\
&\quad
G_{F}=1.1663787\times10^{-5}{\rm GeV^{-2}}.
\end{split}
\end{eqnarray}

For the $B$ meson wave function, we adopt the model
\begin{equation}
\phi_{B}(x, b)=N_{B} x^{2}(1-x)^{2} \exp \left[-\frac{M_{B}^{2} x^{2}}{2 \omega_{b}^{2}}-\frac{1}{2}\left(\omega_{b} b\right)^{2}\right],
\end{equation}
where $\omega_{b}$ is a free parameter  and we take $\omega_{b}=0.4\pm 0.04 \rm GeV$ and $N_{B}=91.7456$ is the normalization factor for $\omega_{b}=0.4 \rm GeV$ \cite{YK2001,LKM}. This is the best fit for most of the measured hadronic $B$ decays. For the light meson wave function, we neglect the $b$ dependent part, which is not important in numerical analysis.
We choose the wave function of $\rho (\omega,\phi)$ meson similar to the pion case
for $\phi_{\rho(\omega,\phi)}$, $\phi_{\rho(\omega,\phi)}^{t}$, and $\phi_{\rho(\omega,\phi)}^{s}$ \cite{PBALL29,bib30,bib31}. The relevant Gegenbauer polynomials are defined by
$C_{2}^{3 / 2}(t)=\frac{3}{2}\left(5 t^{2}-1\right)$,
$C_{4}^{1 / 2}(t)=\frac{1}{8}\left(35 t^{4}-30 t^{2}+3\right)$ \cite{PBALL29}.
The two input parameters $f_{q}$ and $f_{s}$, in the quark-flavor basis have been extracted from various related experiments \cite{bib20,bib21}.
The other parameters can be found in  \cite{etawavefunction,paper11,PBALL,PBALL27,PBALL28}.

\subsection{\label{num}Numerical results}
\begin{figure}[h]
\centering
\includegraphics[height=5cm,width=7.5cm]{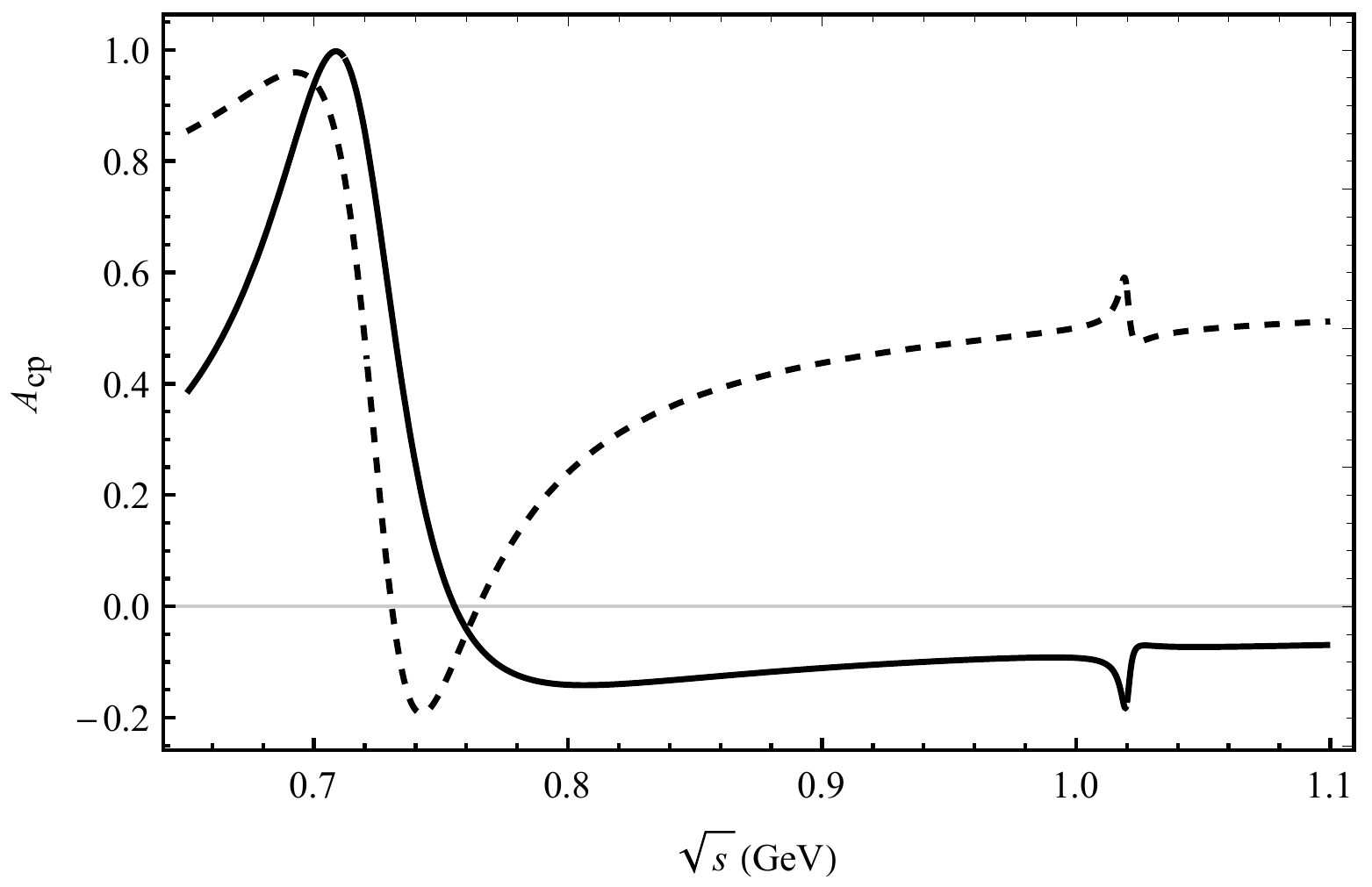}
\caption{Plot of  $A_{CP}$ as a function of $\sqrt{s}$ corresponding to central parameter values of CKM matrix elements.
The solid (dashed) line corresponds to the decay channel of $B^{0}\rightarrow\pi^+\pi^{-}\eta$
($B^{0}\rightarrow\pi^+\pi^{-}\eta'$).}
\label{fig1}
\end{figure}

\begin{figure}[h]
\centering
\includegraphics[height=5cm,width=7.5cm]{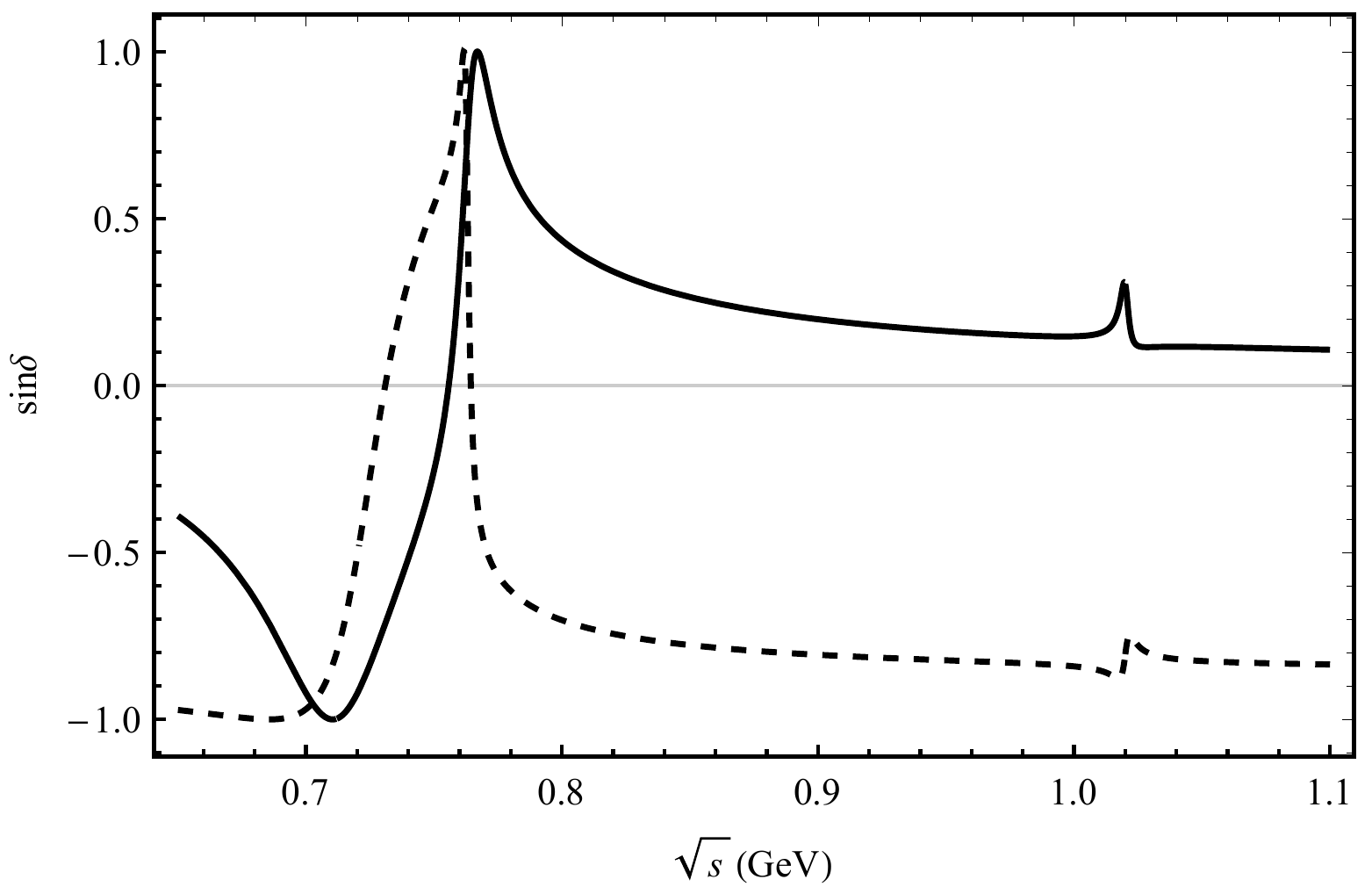}
\caption{Plot of ${\rm{sin}}\delta$ as a function of $\sqrt{s}$ corresponding to central parameter values of CKM matrix elements.
The solid (dashed) line corresponds to the decay channel of $B^{0}\rightarrow\pi^+\pi^{-}\eta$
($B^{0}\rightarrow\pi^+\pi^{-}\eta'$).}
\label{fig2}
\end{figure}

\begin{figure}[h]
\centering
\includegraphics[height=5cm,width=7.5cm]{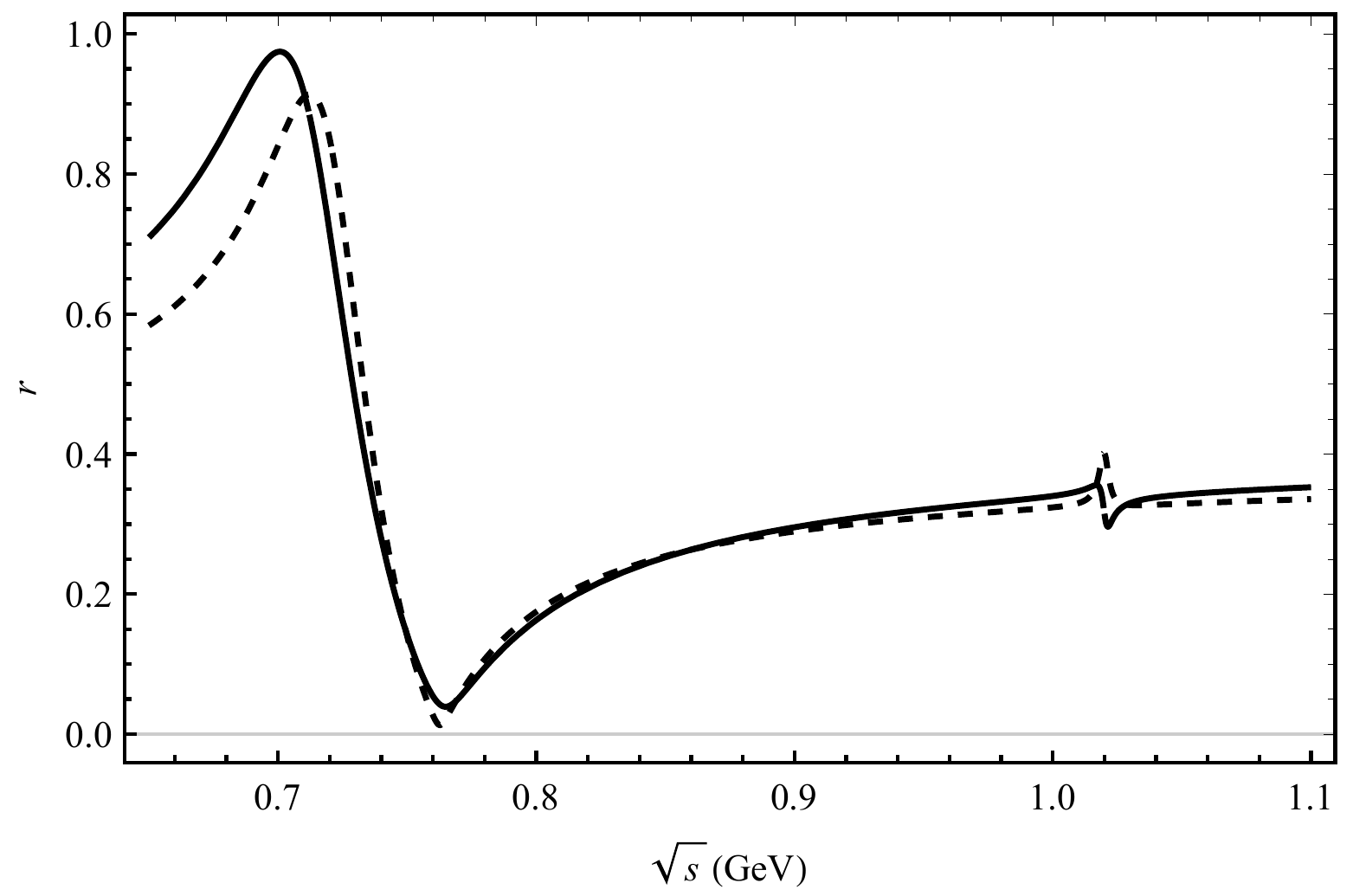}
\caption{Plot of $r$ as a function of $\sqrt{s}$ corresponding to central parameter values of CKM matrix elements.
The solid (dashed) line corresponds to the decay channel of $B^{0}\rightarrow\pi^+\pi^{-}\eta$
($B^{0}\rightarrow\pi^+\pi^{-}\eta'$).}
\label{fig3}
\end{figure}

In the framework of perturbative QCD, we find that the $CP$ violation
is changed sharply for the decay processes of ${B}^{0}\rightarrow\pi^+\pi^{-}\eta$ and
${B}^{0}\rightarrow\pi^+\pi^{-}\eta'$ from the  $\rho-\omega-\phi$ resonance in
the vicinity of $\omega$ and $\phi$ mass.
The results are shown in Fig.\ref{fig1}, Fig.\ref{fig2} and Fig.\ref{fig3}, respectively.
The plot of the $CP$ violation as a function of $\sqrt{s}$ is presented in Fig.\ref{fig1}.
One can find the $CP$ violation varies sharply
when the invariant masses of the $\pi^+\pi^{-}$ pairs are in the area around the $\omega$ resonance range and changes slightly around the $\phi$ resonance range.
For the decay channel of $B^{0}\rightarrow\pi^+\pi^{-}\eta$, we obtain the $CP$ violation varies
from $99.6\%$ to $-14.2\%$ and from $-18.8\%$ to $-6.3\%$ at the $\rho-\omega$ resonance range and the $\rho-\phi$ resonance range, respectively.
For the decay channel of $B^{0}\rightarrow\pi^+\pi^{-}\eta'$, the $CP$ violation varies
from $95.9\%$ to $-18.8\%$ and from $47.1\%$ to $59.5\%$ at the $\rho-\omega$ resonance range and the $\rho-\phi$ resonance range, respectively.

 One can find the $CP$ violation is affected by the weak phase difference, the strong phase difference and $r$. The weak phase depends on the CKM matrix elements which have little effect on our results.
Hence, we present the results corresponding to central parameter values of CKM matrix elements.
In Fig.\ref{fig2}, we present the plot of ${\rm{sin}}\delta$ as a function $\sqrt{s}$ from the central parameter values $\rho$, $\eta$, $\lambda$ and $A$ of CKM matrix elements. One can see the ${\rm{sin}}\delta$ oscillates considerably at the area of $\omega$ resonance, and changes slightly at the area of $\phi$ resonance.
The plot of $r$ as a function of $\sqrt{s}$ is presented in Fig.\ref{fig3}.
The $r$ changes sharply for the $\omega$ resonance range and slightly for the $\phi$ resonance range.

\section{\label{sum}SUMMARY AND DISCUSSION}

In this paper, we introduce the formalism for the $\rho-\omega-\phi$ mesons interferences from
the isospin symmetry breaking. The new strong phase can be produced by the resonance contributions
of $\rho-\omega$, $\rho-\phi$ and $\omega-\phi$. The mechanism is applied to
the decay process of $B^{0}\rightarrow\pi^+\pi^{-}\eta^{(')}$.
It has been found the $CP$ asymmetry oscillates greatly for the
resonance range.
The maximum $CP$ asymmetry can reach $99.6\%$ and $-18.2\%$ in the vicinity of the $\omega$ resonance range and the $\phi$ resonance range for the decay process of $B^{0}\rightarrow\pi^+\pi^{-}\eta$, respectively. For the decay process of $B^{0}\rightarrow\pi^+\pi^{-}\eta'$, the maximum $CP$ asymmetry is $95.9\%$ and $59.5\%$ at the area of $\omega$ resonance and $\phi$ resonance.
Our formalism can be used to calculate the other decay process.

Detection of $CP$ violation signal is an important field in the $B$ meson decay process.
For the three bodies final states, the $CP$ violation is often dominated by quasi-two-body decay channels and depends on the relative phase between the two quasi-two-body amplitudes.
The numbers needed for observing the large $CP$ violation depend
on both the magnitudes of the $CP$ violation
 and the branching ratios of heavy $B$ meson decays.
 We find that the contribution of three meson mixing has little
effect on the branching ratio and can be ignored safely because
the mechanism can only provide the strong phase.
 For one (three) standard deviation signature,
the number of $B\bar{B}$ pairs we need is
{\cite{Du1986,Louis,WT}}
\begin{eqnarray}
N_{B\bar{B}}\sim \frac{1}{BRA_{CP}^{2}}(1-A_{CP}^{2})\Bigg(\frac{9}{BR
A_{CP}^{2}}(1-A_{CP}^{2})\Bigg),
\end{eqnarray}
where BR is the branching ratio for $B\rightarrow \rho^{0}\eta^{(')}$.
We present the numbers of $B\bar{B}$ pairs for observing
the large $CP$ violation at LHC.
For the channel
 $B^{0}\rightarrow \rho^{0}(\omega,\phi)\eta\rightarrow \pi^{+}\pi^{-} \eta$,
the numbers of  $B\bar{B}$ pairs are $10^{4}$ ($10^{5}$) and
$10^{8}$ ($10^{9}$) in the resonance ranges of $\omega$ and $\phi$ for $1\sigma$ ($3\sigma$) signature.
We need $10^{5}$ ($10^{6}$) and $10^{7}$ ($10^{8}$) $B\bar{B}$ pairs to observe the
$CP$ violation from the two resonance ranges in the decay process of
 $B^{0}\rightarrow \rho^{0}(\omega,\phi)\eta^{'}\rightarrow \pi^{+}\pi^{-} \eta^{'}$
for $1\sigma$ ($3\sigma$) signature, respectively.

The Large Hadron Collider (LHC) worked successfully for a proton-proton collisions with a 7 TeV
center-of-mass energy at CERN in 2010.
With the designed center-of-mass energy $14$ TeV and luminosity $L=10^{34}
cm^{-2}s^{-1}$, the LHC provides new possibilities to search $CP$ violation and new physics.
 The $b\bar{b}$ production cross section
will be of the order 0.5 mb, providing as many as $0.5\times
10^{12}$ bottom events per year at the LHC {\cite{Schopper2005,cern2018}}.
The LHC can provide about $10^{13}$ $B\bar{B}$ paires.
In particular, the LHCb detector is designed to study
$CP$ violation and  rare decays in $b$-hadron systems precisely
by large number of $b$-hadrons produced at the LHC.
Direct $CP$ violation can be observed in the decay processes $B$ and $\bar{B}$ for
the differences from the LHCb detector.
At the same time, the ATLAS and CMS experiments are expected for discovering new
physics and focus on most of their $B$ physics program within
the first few years {\cite{Schopper2005,cern2018}}.
To extend its discovery potential, the LHC
made a major upgrade and increased its luminosity by a factor of
five beyond its design value recently.
 Hence, it is very
possible to observe the large $CP$ violation in small
energy range of $\rho^{0}\sim \omega$  and  $\rho^{0}\sim \phi$ resonances at the peak values of $CP$ violation
from the LHC experiment due to the high luminosity large hadron collider (HL-LHC)
even though the branching fractions in these regions may be tiny.
For the experiments, it is possible to reconstruction $\pi^{+}$, $\pi^{-}$ and $\eta^{(')}$ mesons
when the invariant
masses of $\pi^{+}\pi^{-}$ pairs are in the vicinity of the $\omega$ or $\phi$
resonances. Therefore, it is very
possible to observe the large $CP$ violation in
$B^{0}\rightarrow \rho^{0}(\omega,\phi)\eta^{(')}\rightarrow \pi^{+}\pi^{-} \eta^{(')}$ at the LHC.

\section{Acknowledgments}
This work was supported by National Natural Science
Foundation of China (Project Numbers 11605041).

\section{APPENDIX: Related functions defined in the text}
The functions related with the tree and penguin contributions are presented  with PQCD approach \cite{LKM,AMLi2007,YK2001}.

The hard scales $t$ are chosen as \begin{eqnarray}
%t_a&=&\mbox{max}\{{\sqrt{x_2}m_{B},1/b_2,1/b_3}\},\\
t_e^1&=&\mbox{max}\{{\sqrt{x_3}m_{B},1/b_1,1/b_3}\},\\
t_e^2&=&\mbox{max}\{{\sqrt{x_1}m_{B},1/b_1,1/b_3}\},\\
%t_e^3&=&\mbox{max}\{\sqrt{x_3}m_{B},1/b_2,1/b_3\},\\
%t_e^4&=&\mbox{max}\{\sqrt{x_2}m_{B},1/b_2,1/b_3\},\\
t_f&=&\mbox{max}\{\sqrt{x_1x_3}m_{B},\sqrt{(x_1-x_2)x_3}m_{B},1/b_1,1/b_2\},\\
t_f^1&=&\mbox{max}\{{\sqrt{x_2x_3}m_{B},1/b_1,1/b_2}\},\\
t_f^2&=&\mbox{max}\{{\sqrt{x_2x_3}m_{B},\sqrt{x_2+x_3-x_2x_3}m_{B},1/b_1,1/b_2}\},\\
t_f^3&=&\mbox{max}\{\sqrt{x_1+x_2+x_3-x_1x_3-x_2x_3}m_{B},\sqrt{x_2x_3}m_{B},1/b_1,1/b_2\},\\
t_f^4&=&\mbox{max}\{\sqrt{x_2x_3}m_{B},\sqrt{(x_1-x_2)x_3}m_{B},1/b_1,1/b_2\}.
\end{eqnarray}

The function $h$ coming from the Fourier transformations of the function $H^{(0)}$ \cite{xiao2007}.
They are defined by

\begin{equation}
\begin{aligned}
h_{e}\left(x_{1}, x_{3}, b_{1}, b_{3}\right)=& K_{0}\left(\sqrt { x _ { 1 } x _ { 3 } } m _ { B } b _ { 1 } \left[\theta\left(b_{1}-b_{3}\right) K_{0}\left(\sqrt{x}_{3} m_{B} b_{1}\right) I_{0}\left(\sqrt{x}_{3} m_{B} b_{3}\right)\right.\right.\\
&\left.+\theta\left(b_{3}-b_{1}\right) K_{0}\left(\sqrt{x}_{3} m_{B} b_{3}\right) I_{0}\left(\sqrt{x}_{3} m_{B} b_{1}\right)\right] S_{t}\left(x_{3}\right),
\end{aligned}
\end{equation}

\begin{equation}
\begin{aligned}
h_{e}^{1}\left(x_{1}, x_{2}, b_{1}, b_{2}\right)=& K_{0}\left(\sqrt { x _ { 1 } x _ { 2 } } m _ { B } b _ { 1 } \left[\theta\left(b_{1}-b_{2}\right) K_{0}\left(\sqrt{x}_{2} m_{B} b_{1}\right) I_{0}\left(\sqrt{x}_{2} m_{B} b_{2}\right)\right.\right.\\
&\left.+\theta\left(b_{2}-b_{1}\right) K_{0}\left(\sqrt{x}_{2} m_{B} b_{2}\right) I_{0}\left(\sqrt{x}_{2} m_{B} b_{1}\right)\right],
\end{aligned}
\end{equation}

\begin{equation}
\begin{aligned}
h_{f}\left(x_{1}, x_{2}, x_{3}, b_{1}, b_{2}\right)=&\left[\theta\left(b_{2}-b_{1}\right) I_{0}\left(M_{B} \sqrt{x_{1} x_{3}} b_{1}\right) K_{0}\left(M_{B} \sqrt{x_{1} x_{3}} b_{2}\right)\right.\\
&\left.+\left(b_{1} \longleftrightarrow b_{2}\right)\right] \cdot\left\{\begin{array}{ll}
K_{0}\left(M_{B} F_{(1)} b_{2}\right), & \text { for } F_{(1)}^{2}>0 \\
\frac{i \pi}{2} H_{0}^{(1)}\left(M_{B} \sqrt{\left.\left|F_{(1)^{2}}\right| b_{2}\right),}\right. & \text { for } F_{(1)}^{2}<0
\end{array},\right.
\end{aligned}
\end{equation}

\begin{equation}
\begin{aligned}
h_{f}^{1}\left(x_{1}, x_{2}, x_{3}, b_{1}, b_{2}\right)=& K_{0}\left(-i \sqrt{x_{2} x_{3}} m_{B} b_{1}\left[\theta\left(b_{1}-b_{2}\right) K_{0}\left(-i \sqrt{x}_{2} x_{3} m_{B} b_{1}\right) J_{0}\left(\sqrt{x}_{2} x_{3} m_{B} b_{2}\right)\right.\right.\\
&\left.+\theta\left(b_{2}-b_{1}\right) K_{0}\left(-i \sqrt{x}_{2} x_{3} m_{B} b_{2}\right) J_{0}\left(\sqrt{x}_{2} x_{3} m_{B} b_{3}\right)\right],
\end{aligned}
\end{equation}

\begin{equation}
\begin{aligned}
h_{f}^{2}\left(x_{1}, x_{2}, x_{3}, b_{1}, b_{2}\right)=& K_{0}\left(i \sqrt { x _ { 2 } + x _ { 3 } - x _ { 2 } x _ { 3 } } m _ { B } b _ { 1 } \left[\theta\left(b_{1}-b_{2}\right) K_{0}\left(-i \sqrt{x}_{2} x_{3} m_{B} b_{1}\right) J_{0}\left(\sqrt{x}_{2} x_{3} m_{B} b_{2}\right)\right.\right.\\
&\left.+\theta\left(b_{2}-b_{1}\right) K_{0}\left(-i \sqrt{x}_{2} x_{3} m_{B} b_{2}\right) J_{0}\left(\sqrt{x}_{2} x_{3} m_{B} b_{1}\right)\right],
\end{aligned}
\end{equation}

\begin{equation}
\begin{aligned}
h_{f}^{3}\left(x_{1}, x_{2}, x_{3}, b_{1}, b_{2}\right)=&\left[\theta\left(b_{1}-b_{2}\right) K_{0}\left(i \sqrt{x_{2} x_{3}} b_{1} M_{B}\right) I_{0}\left(i \sqrt{x_{2} x_{3}} b_{2} M_{B}\right)\right.\\
&\left.+\left(b_{1} \longleftrightarrow b_{2}\right)\right] \cdot K_{0}\left(\sqrt{x_{1}+x_{2}+x_{3}-x_{1} x_{3}-x_{2} x_{3}} b_{1} M_{B}\right),
\end{aligned}
\end{equation}

\begin{equation}
\begin{aligned}
h_{f}^{4}\left(x_{1}, x_{2}, x_{3}, b_{1}, b_{2}\right)=&\left[\theta\left(b_{1}-b_{2}\right) K_{0}\left(i \sqrt{x_{2} x_{3}} b_{1} M_{B}\right) I_{0}\left(i \sqrt{x_{2} x_{3}} b_{2} M_{B}\right)\right.\\
&\left.+\left(b_{1} \longleftrightarrow b_{2}\right)\right] \cdot\left\{\begin{array}{ll}
K_{0}\left(M_{B} F_{(2)} b_{2}\right), & \text { for } F_{(2)}^{2}>0 \\
\frac{i \pi}{2} H_{0}^{(1)}\left(M_{B} \sqrt{\left|F_{(2)^{2}}\right|} b_{2}\right), & \text { for } F_{(2)}^{2}<0
\end{array},\right.
\end{aligned}
\end{equation}

where $J_0$ is the Bessel function and $K_0$, $I_0$ are the modified Bessel functions $K_0(-ix) = -\frac{\pi}{2}\mathrm{y}_0(x) + i\,\frac{\pi}{2} \mathrm{J}_0(x)$, and $F_{(j)}$'s are defined by
\begin{eqnarray}
F_{(1)}^2=(x_1-x_2)x_3, \;\;\;\;  F_{(2)}^2=(x_1-x_2)x_3.
\end{eqnarray}

The $S_t$ re-sums the threshold logarithms $\ln^2x$ appearing in the
hard kernels to all orders and it has been parameterized as
\begin{eqnarray}
S_t(x)=\frac{2^{1+2c}\Gamma(3/2+c)}{\sqrt \pi
\Gamma(1+c)}[x(1-x)]^c,
\end{eqnarray}
with $c=0.3$. In the nonfactorizable contributions, $S_t(x)$ gives
a very small numerical effect on the amplitude~\cite{L4}.

The Sudakov exponents are defined as
\begin{equation}
S_{a b}(t)=s\left(x_{1} \frac{m_{B}}{\sqrt{2}}, b_{1}\right)+s\left(x_{3} \frac{m_{B}}{\sqrt{2}}, b_{3}\right)+s\left(\left(1-x_{3}\right) \frac{m_{B}}{\sqrt{2}}, b_{3}\right)-\frac{1}{\beta_{1}}\left[\ln \frac{\ln (t / \Lambda)}{-\ln \left(b_{1} \Lambda\right)}+\ln \frac{\ln (t / \Lambda)}{-\ln \left(b_{3} \Lambda\right)}\right],
\end{equation}

\begin{equation}
\begin{aligned}
S_{c d}(t)=& s\left(x_{1} \frac{m_{B}}{\sqrt{2}}, b_{1}\right)+s\left(x_{2} \frac{m_{B}}{\sqrt{2}}, b_{2}\right)+s\left(\left(1-x_{2}\right) \frac{m_{B}}{\sqrt{2}}, b_{2}\right)+s\left(x_{3} \frac{m_{B}}{\sqrt{2}}, b_{1}\right)+s\left(\left(1-x_{3}\right) \frac{m_{B}}{\sqrt{2}}, b_{1}\right) \\
&-\frac{1}{\beta_{1}}\left[\ln \frac{\ln (t / \Lambda)}{-\ln \left(b_{1} \Lambda\right)}+\ln \frac{\ln (t / \Lambda)}{-\ln \left(b_{2} \Lambda\right)}\right],
\end{aligned}
\end{equation}

\begin{equation}
\begin{aligned}
S_{e f}(t)=& s\left(x_{1} \frac{m_{B}}{\sqrt{2}}, b_{1}\right)+s\left(x_{2} \frac{m_{B}}{\sqrt{2}}, b_{2}\right)+s\left(\left(1-x_{2}\right) \frac{m_{B}}{\sqrt{2}}, b_{2}\right)+s\left(x_{3} \frac{m_{B}}{\sqrt{2}}, b_{2}\right)+s\left(\left(1-x_{3}\right) \frac{m_{B}}{\sqrt{2}}, b_{2}\right) \\
&-\frac{1}{\beta_{1}}\left[\ln \frac{\ln (t / \Lambda)}{-\ln \left(b_{1} \Lambda\right)}+2 \ln \frac{\ln (t / \Lambda)}{-\ln \left(b_{2} \Lambda\right)}\right].
\end{aligned}
\end{equation}

The explicit form for the  function
$s(k,b)$ is \cite{LKM}:
\begin{equation}
\begin{aligned}
s(k, b)=& \frac{2}{3 \beta_{1}}\left[\hat{q} \ln \left(\frac{\hat{q}}{\hat{b}}-\hat{q}+\hat{b}\right)\right]+\frac{A^{(2)}}{4 \beta_{1}^{2}}\left(\frac{\hat{q}}{\hat{b}}-1\right) \\
&-\left[\frac{A^{(2)}}{4 \beta_{1}^{2}}-\frac{1}{3 \beta_{1}}\left(2 \gamma_{E}-1-\ln 2\right)\right] \ln \left(\frac{\hat{q}}{\hat{b}}\right),
\end{aligned}
\end{equation}
where the variables are defined by
\begin{eqnarray}
\hat q\equiv \mbox{ln}[k/(\sqrt \Lambda)],~~~ \hat b\equiv
\mbox{ln}[1/(b\Lambda)], \end{eqnarray}
and the coefficients
$A^{(i)}$ and $\beta_i$ are
\begin{eqnarray}
A^{(2)}&=&\frac{67}{9}
-\frac{\pi^2}{3}-\frac{10}{27}n_f+\frac{8}{3}\beta_1\mbox{ln}(\frac{1}{2}e^{\gamma_E}),\\
\beta_1&=&\frac{33-2n_f}{12},
\end{eqnarray}
where $n_f$ is the number of the quark flavors and $\gamma_E$ is the Euler constant.

The decay amplitude $F_{e}$, $F_{e \rho}$ and $F_{e \omega}$ induced by inserting the $(V-A)(V-A)$ operators are\cite{liuxin}
\begin{equation}
\begin{aligned}
F_{e}=& 4 \sqrt{2} \pi G_{F} C_{F} f_{\rho} m_{B}^{4} \int_{0}^{1} d x_{1} d x_{3} \int_{0}^{\infty} b_{1} d b_{1} b_{3} d b_{3} \phi_{B}\left(x_{1}, b_{1}\right) \cdot\left\{\left[\left(1+x_{3}\right) \phi_{\eta}^{A}\left(x_{3}, b_{3}\right)+r_{\eta}\left(1-2 x_{3}\right)\left(\phi_{\eta}^{P}\left(x_{3}, b_{3}\right)\right.\right.\right.\\
&\left.\left.\left.+\phi_{\eta}^{T}\left(x_{3}, b_{3}\right)\right)\right] \cdot \alpha_{s}\left(t_{e}^{1}\right) h_{e}\left(x_{1}, x_{3}, b_{1}, b_{3}\right) \exp \left[-S_{a b}\left(t_{e}^{1}\right)\right]+2 r_{\eta} \phi_{\eta}^{P}\left(x_{3}, b_{3}\right) \alpha_{s}\left(t_{e}^{2}\right) h_{e}\left(x_{3}, x_{1}, b_{3}, b_{1}\right) \exp \left[-S_{a b}\left(t_{e}^{2}\right)\right]\right\},
\end{aligned}
\end{equation}

\begin{equation}
\begin{aligned}
F_{e \rho}=& 4 \sqrt{2} G_{F} \pi C_{F} m_{B}^{4} \int_{0}^{1} d x_{1} d x_{3} \int_{0}^{\infty} b_{1} d b_{1} b_{3} d b_{3} \phi_{B}\left(x_{1}, b_{1}\right) \cdot\left\{\left[\left(1+x_{3}\right) \phi_{\rho}\left(x_{3}, b_{3}\right)+\left(1-2 x_{3}\right) r_{\rho}\left(\phi_{\rho}^{s}\left(x_{3}, b_{3}\right)\right.\right.\right.\\
&\left.\left.\left.+\phi_{\rho}^{t}\left(x_{3}, b_{3}\right)\right)\right] \alpha_{s}\left(t_{e}^{1}\right) h_{e}\left(x_{1}, x_{3}, b_{1}, b_{3}\right) \exp \left[-S_{a b}\left(t_{e}^{1}\right)\right]+2 r_{\rho} \phi_{\rho}^{s}\left(x_{3}, b_{3}\right) \alpha_{s}\left(t_{e}^{2}\right) h_{e}\left(x_{3}, x_{1}, b_{3}, b_{1}\right) \exp \left[-S_{a b}\left(t_{e}^{2}\right)\right]\right\},
\end{aligned}
\end{equation}

\begin{equation}
\begin{aligned}
F_{e \omega}=& 4 \sqrt{2} G_{F} \pi C_{F} m_{B}^{4} \int_{0}^{1} d x_{1} d x_{3} \int_{0}^{\infty} b_{1} d b_{1} b_{3} d b_{3} \phi_{B}\left(x_{1}, b_{1}\right)\left\{\left[\left(1+x_{3}\right) \phi_{\omega}\left(x_{3}, b_{3}\right)+\left(1-2 x_{3}\right) r_{\omega}\left(\phi_{\omega}^{s}\left(x_{3}, b_{3}\right)\right.\right.\right.\\
&\left.\left.\left.+\phi_{\omega}^{t}\left(x_{3}, b_{3}\right)\right)\right] \alpha_{s}\left(t_{e}^{1}\right) h_{e}\left(x_{1}, x_{3}, b_{1}, b_{3}\right) \exp \left[-S_{a b}\left(t_{e}^{1}\right)\right]+2 r_{\omega} \phi_{\omega}^{s}\left(x_{3}, b_{3}\right) \alpha_{s}\left(t_{e}^{2}\right) h_{e}\left(x_{3}, x_{1}, b_{3}, b_{1}\right) \exp \left[-S_{a b}\left(t_{e}^{2}\right)\right]\right\}.
\end{aligned}
\end{equation}

From the $(S+P)(S-P)$ operators, we can get
\begin{equation}
\begin{aligned}
F_{e \rho}^{P}=& 8 \sqrt{2} G_{F} \pi C_{F} f_{\eta}^{d} m_{B}^{4} \int_{0}^{1} d x_{1} d x_{3} \int_{0}^{\infty} b_{1} d b_{1} b_{3} d b_{3} \phi_{B}\left(x_{1}, b_{1}\right) \cdot\left\{\left[\phi_{\rho}\left(x_{3}, b_{3}\right)+r_{\rho}\left(\left(x_{3}+2\right) \phi_{\rho}^{s}\left(x_{3}, b_{3}\right)-x_{3} \phi_{\rho}^{t}\left(x_{3}, b_{3}\right)\right)\right]\right.\\
&\left.\cdot \alpha_{s}\left(t_{e}^{1}\right) h_{e}\left(x_{1}, x_{3}, b_{1}, b_{3}\right) \exp \left[-S_{a b}\left(t_{e}^{1}\right)\right]+\left(x_{1} \phi_{\rho}\left(x_{3}, b_{3}\right)+2 r_{\rho} \phi_{\rho}^{s}\left(x_{3}, b_{3}\right)\right) \alpha_{s}\left(t_{e}^{2}\right) h_{e}\left(x_{3}, x_{1}, b_{3}, b_{1}\right) \exp \left[-S_{a b}\left(t_{e}^{2}\right)\right]\right\},
\end{aligned}
\end{equation}

\begin{equation}
\begin{aligned}
F_{e \omega}^{P 2}=& 8 \sqrt{2} G_{F} \pi C_{F} r_{\eta} m_{B}^{4} \int_{0}^{1} d x_{1} d x_{3} \int_{0}^{\infty} b_{1} d b_{1} b_{3} d b_{3} \phi_{B}\left(x_{1}, b_{1}\right)\left\{\left[\phi_{\omega}\left(x_{3}, b_{3}\right)+r_{\omega}\left(\left(x_{3}+2\right) \phi_{\omega}^{s}\left(x_{3}, b_{3}\right)-x_{3} \phi_{\omega}^{t}\left(x_{3}, b_{3}\right)\right)\right]\right.\\
&\left.\times \alpha_{s}\left(t_{e}^{1}\right) h_{e}\left(x_{1}, x_{3}, b_{1}, b_{3}\right) \exp \left[-S_{a b}\left(t_{e}^{1}\right)\right]+\left(x_{1} \phi_{\omega}\left(x_{3}, b_{3}\right)+2 r_{\omega} \phi_{\omega}^{s}\left(x_{3}, b_{3}\right)\right) \alpha_{s}\left(t_{e}^{2}\right) h_{e}\left(x_{3}, x_{1}, b_{3}, b_{1}\right) \exp \left[-S_{a b}\left(t_{e}^{2}\right)\right]\right\}.
\end{aligned}
\end{equation}

The decay amplitude for $(V-A)(V-A)$ and $(V-A)(+A)$ operators can be
written as follows
\begin{equation}
\begin{aligned}
M_{e}=&-M_{e}^{P 2}=\frac{16}{\sqrt{3}} G_{F} \pi C_{F} m_{B}^{4} \int_{0}^{1} d x_{1} d x_{2} d x_{3} \int_{0}^{\infty} b_{1} d b_{1} b_{2} d b_{2} \phi_{B}\left(x_{1}, b_{1}\right) \phi_{\omega}\left(x_{2}, b_{2}\right) \\
& \times\left\{\left[2 x_{3} r_{\eta} \phi_{\eta}^{T}\left(x_{3}, b_{1}\right)-x_{3} \phi_{\eta}^{A}\left(x_{3}, b_{1}\right)\right] \alpha_{s}\left(t_{f}\right) h_{f}\left(x_{1}, x_{2}, x_{3}, b_{1}, b_{2}\right) \exp \left[-S_{c d}\left(t_{f}\right)\right]\right\},
\end{aligned}
\end{equation}

\begin{equation}
\begin{aligned}
M_{e \rho}=&-\frac{16}{\sqrt{3}} G_{F} \pi C_{F} m_{B}^{4} \int_{0}^{1} d x_{1} d x_{2} d x_{3} \int_{0}^{\infty} b_{1} d b_{1} b_{2} d b_{2} \phi_{B}\left(x_{1}, b_{1}\right) \phi_{\eta}^{A}\left(x_{2}, b_{2}\right) \\
& \cdot\left\{x_{3}\left[\phi_{\rho}\left(x_{3}, b_{2}\right)-2 r_{\rho} \phi_{\rho}^{t}\left(x_{3}, b_{3}\right)\right] \alpha_{s}\left(t_{f}\right) h_{f}\left(x_{1}, x_{2}, x_{3}, b_{1}, b_{2}\right) \exp \left[-S_{c d}\left(t_{f}\right)\right]\right\},
\end{aligned}
\end{equation}

\begin{equation}
\begin{aligned}
M_{e \omega}=&-\frac{16}{\sqrt{3}} G_{F} \pi C_{F} m_{B}^{4} \int_{0}^{1} d x_{1} d x_{2} d x_{3} \int_{0}^{\infty} b_{1} d b_{1} b_{2} d b_{2} \phi_{B}\left(x_{1}, b_{1}\right) \phi_{\eta}^{A}\left(x_{2}, b_{2}\right) \\
& \times\left\{x_{3}\left[\phi_{\omega}\left(x_{3}, b_{1}\right)-2 r_{\omega} \phi_{\omega}^{t}\left(x_{3}, b_{1}\right)\right] \alpha_{s}\left(t_{f}\right) h_{f}\left(x_{1}, x_{2}, x_{3}, b_{1}, b_{2}\right) \exp \left[-S_{c d}\left(t_{f}\right)\right]\right\},
\end{aligned}
\end{equation}

\begin{equation}
\begin{aligned}
M_{a}=& \frac{16}{\sqrt{3}} \pi G_{F} C_{F} m_{B}^{4} \int_{0}^{1} d x_{1} d x_{2} d x_{3} \int_{0}^{\infty} b_{1} d b_{1} b_{2} d b_{2} \phi_{B}\left(x_{1}, b_{1}\right) \\
& \times\left\{\left[r_{\omega} r_{\eta}\left(x_{3}-x_{2}\right)\left[\phi_{\eta}^{P}\left(x_{3}, b_{2}\right) \phi_{\omega}^{t}\left(x_{2}, b_{2}\right)+\phi_{\eta}^{T}\left(x_{3}, b_{2}\right) \phi_{\omega}^{s}\left(x_{2}, b_{2}\right)\right]+r_{\omega} r_{\eta}\left(x_{2}+x_{3}\right)\right.\right.\\
&\left.\times\left[\phi_{\eta}^{P}\left(x_{3}, b_{2}\right) \phi_{\omega}^{s}\left(x_{2}, b_{2}\right)+\phi_{\eta}^{T}\left(x_{3}, b_{2}\right) \phi_{\omega}^{t}\left(x_{2}, b_{2}\right)\right]+x_{3} \phi_{\omega}\left(x_{2}, b_{2}\right) \phi_{\eta}^{A}\left(x_{3}, b_{2}\right)\right] \alpha_{s}\left(t_{f}^{4}\right) h_{f}^{4}\left(x_{1}, x_{2}, x_{3}, b_{1}, b_{2}\right) \\
& \times \exp \left[-S_{e f}\left(t_{f}^{4}\right)\right]-\left[r_{\omega} r_{\eta}\left(x_{2}-x_{3}\right)\left[\phi_{\eta}^{P}\left(x_{3}, b_{2}\right) \phi_{\omega}^{t}\left(x_{2}, b_{2}\right)+\phi_{\eta}^{T}\left(x_{3}, b_{2}\right) \phi_{\omega}^{s}\left(x_{2}, b_{2}\right)\right]\right.\\
&\left.+r_{\omega} r_{\eta}\left[\left(2+x_{2}+x_{3}\right) \phi_{\eta}^{P}\left(x_{3}, b_{2}\right) \phi_{\omega}^{s}\left(x_{2}, b_{2}\right)-\left(2-x_{2}-x_{3}\right) \phi_{\eta}^{T}\left(x_{3}, b_{2}\right) \phi_{\omega}^{t}\left(x_{2}, b_{2}\right)\right]+x_{2} \phi_{\omega}\left(x_{2}, b_{2}\right) \phi_{\eta}^{A}\left(x_{3}, b_{2}\right)\right] \\
&\left.\times \alpha_{s}\left(t_{f}^{3}\right) h_{f}^{3}\left(x_{1}, x_{2}, x_{3}, b_{1}, b_{2}\right) \exp \left[-S_{e f}\left(t_{f}^{3}\right)\right]\right\},
\end{aligned}
\end{equation}

\begin{equation}
\begin{aligned}
M_{a \rho}=& \frac{16}{\sqrt{3}} G_{F} \pi C_{F} m_{B}^{4} \int_{0}^{1} d x_{1} d x_{2} d x_{3} \int_{0}^{\infty} b_{1} d b_{1} b_{2} d b_{2} \phi_{B}\left(x_{1}, b_{1}\right) \cdot\left\{\left[x_{3} \phi_{\rho}\left(x_{3}, b_{2}\right) \phi_{\eta}^{A}\left(x_{2}, b_{2}\right)\right.\right.\\
&+r_{\rho} r_{\eta}\left(\left(x_{3}-x_{2}\right)\left(\phi_{\eta}^{P}\left(x_{2}, b_{2}\right) \phi_{\rho}^{t}\left(x_{3}, b_{2}\right)+\phi_{\eta}^{T}\left(x_{2}, b_{2}\right) \cdot \phi_{\rho}^{s}\left(x_{3}, b_{2}\right)\right)+\left(x_{3}+x_{2}\right)\left(\phi_{\eta}^{P}\left(x_{2}, b_{2}\right) \phi_{\rho}^{s}\left(x_{3}, b_{2}\right)\right.\right.\\
&\left.\left.\left.+\phi_{\eta}^{T}\left(x_{2}, b_{2}\right) \phi_{\rho}^{t}\left(x_{3}, b_{2}\right)\right)\right)\right] \cdot \alpha_{s}\left(t_{f}^{1}\right) h_{f}^{1}\left(x_{1}, x_{2}, x_{3}, b_{1}, b_{2}\right) \exp \left[-S_{e f}\left(t_{f}^{1}\right)\right]-\left[x_{2} \phi_{\rho}\left(x_{3}, b_{2}\right) \phi_{\eta}^{A}\left(x_{2}, b_{2}\right)\right.\\
&+r_{\rho} r_{\eta}\left(\left(x_{2}-x_{3}\right)\left(\phi_{\eta}^{P}\left(x_{2}, b_{2}\right) \phi_{\rho}^{t}\left(x_{3}, b_{2}\right)+\phi_{\eta}^{T}\left(x_{2}, b_{2}\right) \phi_{\rho}^{s}\left(x_{3}, b_{2}\right)\right)+r_{\rho} r_{\eta} \cdot\left(\left(2+x_{2}+x_{3}\right) \phi_{\eta}^{P}\left(x_{2}, b_{2}\right) \phi_{\rho}^{s}\left(x_{3}, b_{2}\right)\right.\right.\\
&\left.\left.\left.\left.-\left(2-x_{2}-x_{3}\right) \phi_{\eta}^{T}\left(x_{2}, b_{2}\right) \phi_{\rho}^{t}\left(x_{3}, b_{2}\right)\right)\right)\right] \cdot \alpha_{s}\left(t_{f}^{2}\right) h_{f}^{2}\left(x_{1}, x_{2}, x_{3}, b_{1}, b_{2}\right) \exp \left[-S_{e f}\left(t_{f}^{2}\right)\right]\right\},
\end{aligned}
\end{equation}

where $r_{\eta}\equiv r_{\pi} = m^{\pi}_{0}/m_{B}$.

\begin{equation}
\begin{aligned}
M_{a \omega}=& \frac{16}{\sqrt{3}} G_{F} \pi C_{F} m_{B}^{4} \int_{0}^{1} d x_{1} d x_{2} d x_{3} \int_{0}^{\infty} b_{1} d b_{1} b_{2} d b_{2} \phi_{B}\left(x_{1}, b_{1}\right)\left\{\left[x_{3} \phi_{\omega}\left(x_{3}, b_{2}\right) \phi_{\eta}^{A}\left(x_{2}, b_{2}\right)\right.\right.\\
&+r_{\omega} r_{\eta}\left(\left(x_{3}-x_{2}\right)\left(\phi_{\eta}^{P}\left(x_{2}, b_{2}\right) \phi_{\omega}^{t}\left(x_{3}, b_{2}\right)+\phi_{\eta}^{T}\left(x_{2}, b_{2}\right) \phi_{\omega}^{s}\left(x_{3}, b_{2}\right)\right)+\left(x_{3}+x_{2}\right)\left(\phi_{\eta}^{P}\left(x_{2}, b_{2}\right) \phi_{\omega}^{s}\left(x_{3}, b_{2}\right)\right.\right.\\
&\left.\left.+\phi_{\eta}^{T}\left(x_{2}, b_{2}\right) \phi_{\omega}^{t}\left(x_{3}, b_{2}\right)\right)\right] \alpha_{s}\left(t_{f}^{4}\right) h_{f}^{4}\left(x_{1}, x_{2}, x_{3}, b_{1}, b_{2}\right) \exp \left[-S_{e f}\left(t_{f}^{4}\right)\right]-\left[x_{2} \phi_{\omega}\left(x_{3}, b_{2}\right) \phi_{\eta}^{A}\left(x_{2}, b_{2}\right)\right.\\
&+r_{\omega} r_{\eta}\left(\left(x_{2}-x_{3}\right)\left(\phi_{\eta}^{P}\left(x_{2}, b_{2}\right) \phi_{\omega}^{t}\left(x_{3}, b_{2}\right)+\phi_{\eta}^{T}\left(x_{2}, b_{2}\right) \phi_{\omega}^{s}\left(x_{3}, b_{2}\right)\right)+r_{\omega} r_{\eta}\left(\left(2+x_{2}+x_{3}\right) \phi_{\eta}^{P}\left(x_{2}, b_{2}\right) \phi_{\omega}^{s}\left(x_{3}, b_{2}\right)\right.\right.\\
&\left.\left.\left.\left.-\left(2-x_{2}-x_{3}\right) \phi_{\eta}^{T}\left(x_{2}, b_{2}\right) \phi_{\omega}^{t}\left(x_{3}, b_{2}\right)\right)\right)\right] \alpha_{s}\left(t_{f}^{3}\right) h_{f}^{3}\left(x_{1}, x_{2}, x_{3}, b_{1}, b_{2}\right) \exp \left[-S_{e f}\left(t_{f}^{3}\right)\right]\right\},
\end{aligned}
\end{equation}

\begin{equation}
\begin{aligned}
M_{e}^{P}=&-\frac{32}{\sqrt{3}} G_{F} \pi C_{F} r_{\rho} m_{B}^{4} \int_{0}^{1} d x_{1} d x_{2} d x_{3} \int_{0}^{\infty} b_{1} d b_{1} b_{2} d b_{2} \phi_{B}\left(x_{1}, b_{1}\right) \cdot\left\{\left[x_{2} \phi_{\eta}^{A}\left(x_{3}, b_{2}\right)\left(\phi_{\rho}^{s}\left(x_{2}, b_{2}\right)-\phi_{\rho}^{t}\left(x_{2}, b_{2}\right)\right)\right.\right.\\
&+r_{\eta}\left(\left(x_{2}+x_{3}\right)\left(\phi_{\eta}^{P}\left(x_{3}, b_{2}\right) \cdot \phi_{\rho}^{s}\left(x_{2}, b_{2}\right)+\phi_{\eta}^{T}\left(x_{3}, b_{2}\right) \phi_{\rho}^{t}\left(x_{2}, b_{2}\right)\right)+\left(x_{3}-x_{2}\right)\left(\phi_{\eta}^{P}\left(x_{3}, b_{2}\right) \phi_{\rho}^{t}\left(x_{2}, b_{2}\right)\right.\right.\\
&\left.\left.\left.\left.+\phi_{\eta}^{T}\left(x_{3}, b_{2}\right) \phi_{\rho}^{s}\left(x_{2}, b_{2}\right)\right)\right)\right] \alpha_{s}\left(t_{f}\right) h_{f}\left(x_{1}, x_{2}, x_{3}, b_{1}, b_{2}\right) \exp \left[-S_{c d}\left(t_{f}\right)\right]\right\},
\end{aligned}
\end{equation}

\begin{equation}
\begin{aligned}
M_{a \rho}^{P}=&-\frac{16}{\sqrt{3}} G_{F} \pi C_{F} m_{B}^{4} \int_{0}^{1} d x_{1} d x_{2} d x_{3} \int_{0}^{\infty} b_{1} d b_{1} b_{2} d b_{2} \phi_{B}\left(x_{1}, b_{1}\right) \cdot\left\{\left[x_{2} r_{\eta} \phi_{\rho}\left(x_{3}, b_{2}\right)\left(\phi_{\eta}^{P}\left(x_{2}, b_{2}\right)+\phi_{\eta}^{T}\left(x_{2}, b_{2}\right)\right)\right.\right.\\
&\left.-x_{3} r_{\rho}\left(\phi_{\rho}^{s}\left(x_{3}, b_{2}\right)+\phi_{\rho}^{t}\left(x_{3}, b_{2}\right)\right) \cdot \phi_{\eta}^{A}\left(x_{2}, b_{2}\right)\right] \alpha_{s}\left(t_{f}^{1}\right) h_{f}^{1}\left(x_{1}, x_{2}, x_{3}, b_{1}, b_{2}\right) \exp \left[-S_{e f}\left(t_{f}^{1}\right)\right] \\
&+\left[\left(2-x_{2}\right) r_{\eta} \phi_{\rho}\left(x_{3}, b_{2}\right)\left(\phi_{\eta}^{P}\left(x_{2}, b_{2}\right)+\phi_{\eta}^{T}\left(x_{2}, b_{2}\right)\right)-\left(2-x_{3}\right) r_{\rho}\left(\phi_{\rho}^{s}\left(x_{3}, b_{2}\right)+\phi_{\rho}^{t}\left(x_{3}, b_{2}\right)\right) \phi_{\eta}^{A}\left(x_{2}, b_{2}\right)\right] \\
&\left.\times \alpha_{s}\left(t_{f}^{2}\right) h_{f}^{2}\left(x_{1}, x_{2}, x_{3}, b_{1}, b_{2}\right) \exp \left[-S_{e f}\left(t_{f}^{2}\right)\right]\right\},
\end{aligned}
\end{equation}

\begin{equation}
M_{a}^{P} = M_{a \rho}^{P}.
\end{equation}

\begin{equation}
\begin{aligned}
M_{a}^{P 2}=&-\frac{16}{\sqrt{3}} \pi G_{F} C_{F} m_{B}^{4} \int_{0}^{1} d x_{1} d x_{2} d x_{3} \int_{0}^{\infty} b_{1} d b_{1} b_{2} d b_{2} \phi_{B}\left(x_{1}, b_{1}\right)\left\{\left[x_{2} \phi_{\omega}\left(x_{2}, b_{2}\right) \phi_{\eta}^{A}\left(x_{3}, b_{2}\right)\right.\right.\\
&+r_{\omega} r_{\eta}\left(\left(x_{2}-x_{3}\right)\left(\phi_{\eta}^{P}\left(x_{3}, b_{2}\right) \phi_{\omega}^{t}\left(x_{2}, b_{2}\right)+\phi_{\eta}^{T}\left(x_{3}, b_{2}\right) \phi_{\omega}^{s}\left(x_{2}, b_{2}\right)\right)+\left(x_{2}+x_{3}\right)\left(\phi_{\eta}^{P}\left(x_{3}, b_{2}\right) \phi_{\omega}^{s}\left(x_{2}, b_{2}\right)\right.\right.\\
&\left.\left.\left.+\phi_{\eta}^{T}\left(x_{3}, b_{2}\right) \phi_{\omega}^{t}\left(x_{2}, b_{2}\right)\right)\right)\right] \alpha_{s}\left(t_{f}^{4}\right) h_{f}^{4}\left(x_{1}, x_{2}, x_{3}, b_{1}, b_{2}\right) \exp \left[-S_{e f}\left(t_{f}^{4}\right)\right]-\left[x_{3} \phi_{\omega}\left(x_{2}, b_{2}\right) \phi_{\eta}^{A}\left(x_{3}, b_{2}\right)\right.\\
&+r_{\omega} r_{\eta}\left(\left(x_{3}-x_{2}\right)\left(\phi_{\eta}^{P}\left(x_{3}, b_{2}\right) \phi_{\omega}^{t}\left(x_{2}, b_{2}\right)+\phi_{\eta}^{T}\left(x_{3}, b_{2}\right) \phi_{\omega}^{s}\left(x_{2}, b_{2}\right)\right)+\left(2+x_{2}+x_{3}\right) \phi_{\eta}^{P}\left(x_{3}, b_{2}\right) \phi_{\omega}^{s}\left(x_{2}, b_{2}\right)\right.\\
&\left.\left.\left.-\left(2-x_{2}-x_{3}\right) \phi_{\eta}^{T}\left(x_{3}, b_{2}\right) \phi_{\omega}^{t}\left(x_{2}, b_{2}\right)\right)\right] \alpha_{s}\left(t_{f}^{3}\right) h_{f}^{3}\left(x_{1}, x_{2}, x_{3}, b_{1}, b_{2}\right) \exp \left[-S_{e f}\left(t_{f}^{3}\right)\right]\right\},
\end{aligned}
\end{equation}

\begin{equation}
\begin{aligned}
M_{a \omega}^{P 1}=&-\frac{16}{\sqrt{3}} G_{F} \pi C_{F} m_{B}^{4} \int_{0}^{1} d x_{1} d x_{2} d x_{3} \int_{0}^{\infty} b_{1} d b_{1} b_{2} d b_{2} \phi_{B}\left(x_{1}, b_{1}\right)\left\{\left[x_{2} r_{\eta} \phi_{\omega}\left(x_{3}, b_{2}\right)\left(\phi_{\eta}^{P}\left(x_{2}, b_{2}\right)+\phi_{\eta}^{T}\left(x_{2}, b_{2}\right)\right)\right.\right.\\
&\left.-x_{3} r_{\omega}\left(\phi_{\omega}^{s}\left(x_{3}, b_{2}\right)+\phi_{\omega}^{t}\left(x_{3}, b_{2}\right)\right) \phi_{\eta}^{A}\left(x_{2}, b_{2}\right)\right] \alpha_{s}\left(t_{f}^{4}\right) h_{f}^{4}\left(x_{1}, x_{2}, x_{3}, b_{1}, b_{2}\right) \exp \left[-S_{e f}\left(t_{f}^{4}\right)\right] \\
&+\left[\left(2-x_{2}\right) r_{\eta} \phi_{\omega}\left(x_{3}, b_{2}\right)\left(\phi_{\eta}^{P}\left(x_{2}, b_{2}\right)+\phi_{\eta}^{T}\left(x_{2}, b_{2}\right)\right)-\left(2-x_{3}\right) r_{\omega}\left(\phi_{\omega}^{s}\left(x_{3}, b_{2}\right)+\phi_{\omega}^{t}\left(x_{3}, b_{2}\right)\right) \phi_{\eta}^{A}\left(x_{2}, b_{2}\right)\right] \\
&\left.\times \alpha_{s}\left(t_{f}^{3}\right) h_{f}^{3}\left(x_{1}, x_{2}, x_{3}, b_{1}, b_{2}\right) \exp \left[-S_{e f}\left(t_{f}^{3}\right)\right]\right\},
\end{aligned}
\end{equation}

\begin{equation}
\begin{aligned}
M_{a \omega}^{P 2}=&-\frac{16}{\sqrt{3}} \pi G_{F} C_{F} m_{B}^{4} \int_{0}^{1} d x_{1} d x_{2} d x_{3} \int_{0}^{\infty} b_{1} d b_{1} b_{2} d b_{2} \phi_{B}\left(x_{1}, b_{1}\right)\left\{\left[x_{2} \phi_{\omega}\left(x_{3}, b_{2}\right) \phi_{\eta}^{A}\left(x_{2}, b_{2}\right)\right.\right.\\
&+r_{\omega} r_{\eta}\left(\left(x_{2}-x_{3}\right)\left(\phi_{\eta}^{P}\left(x_{2}, b_{2}\right) \phi_{\omega}^{t}\left(x_{3}, b_{2}\right)+\phi_{\eta}^{T}\left(x_{2}, b_{2}\right) \phi_{\omega}^{s}\left(x_{3}, b_{2}\right)\right)+\left(x_{2}+x_{3}\right)\left(\phi_{\eta}^{P}\left(x_{2}, b_{2}\right) \phi_{\omega}^{s}\left(x_{3}, b_{2}\right)\right.\right.\\
&\left.\left.\left.+\phi_{\eta}^{T}\left(x_{2}, b_{2}\right) \phi_{\omega}^{t}\left(x_{3}, b_{2}\right)\right)\right)\right] \alpha_{s}\left(t_{f}^{4}\right) h_{f}^{4}\left(x_{1}, x_{2}, x_{3}, b_{1}, b_{2}\right) \exp \left[-S_{e f}\left(t_{f}^{4}\right)\right]-\left[x_{3} \phi_{\omega}\left(x_{3}, b_{2}\right) \phi_{\eta}^{A}\left(x_{2}, b_{2}\right)\right.\\
&+r_{\omega} r_{\eta}\left(\left(x_{3}-x_{2}\right)\left(\phi_{\eta}^{P}\left(x_{2}, b_{2}\right) \phi_{\omega}^{t}\left(x_{3}, b_{2}\right)+\phi_{\eta}^{T}\left(x_{2}, b_{2}\right) \phi_{\omega}^{s}\left(x_{3}, b_{2}\right)\right)+\left(2+x_{2}+x_{3}\right) \phi_{\eta}^{P}\left(x_{2}, b_{2}\right) \phi_{\omega}^{s}\left(x_{3}, b_{2}\right)\right.\\
&\left.\left.\left.-\left(2-x_{2}-x_{3}\right) \phi_{\eta}^{T}\left(x_{2}, b_{2}\right) \phi_{\omega}^{t}\left(x_{3}, b_{2}\right)\right)\right] \alpha_{s}\left(t_{f}^{3}\right) h_{f}^{3}\left(x_{1}, x_{2}, x_{3}, b_{1}, b_{2}\right) \exp \left[-S_{e f}\left(t_{f}^{3}\right)\right]\right\},
\end{aligned}
\end{equation}

replace the $\omega$ meson by the $\phi$ meson, we can get the $M_{a \phi}$ and $M_{a \phi}^{P 2}$ from $M_{a \omega}$ and $M_{a \omega}^{P 2}$.

\begin{equation}
\begin{aligned}
M_{a}^{P 1}=& \frac{16}{\sqrt{3}} G_{F} \pi C_{F} m_{B}^{4} \int_{0}^{1} d x_{1} d x_{2} d x_{3} \int_{0}^{\infty} b_{1} d b_{1} b_{2} d b_{2} \phi_{B}\left(x_{1}, b_{1}\right)\left\{\left[x_{2} r_{\omega} \phi_{\eta}^{A}\left(x_{3}, b_{2}\right)\left(\phi_{\omega}^{s}\left(x_{2}, b_{2}\right)+\phi_{\omega}^{t}\left(x_{2}, b_{2}\right)\right)\right.\right.\\
&\left.-x_{3} r_{\eta}\left(\phi_{\eta}^{P}\left(x_{3}, b_{2}\right)+\phi_{\eta}^{T}\left(x_{3}, b_{2}\right)\right) \phi_{\omega}\left(x_{2}, b_{2}\right)\right] \alpha_{s}\left(t_{f}^{4}\right) h_{f}^{4}\left(x_{1}, x_{2}, x_{3}, b_{1}, b_{2}\right) \exp \left[-S_{e f}\left(t_{f}^{4}\right)\right] \\
&+\left[\left(2-x_{2}\right) r_{\omega} \phi_{\eta}^{A}\left(x_{3}, b_{2}\right)\left(\phi_{\omega}^{s}\left(x_{2}, b_{2}\right)+\phi_{\omega}^{t}\left(x_{2}, b_{2}\right)\right)-\left(2-x_{3}\right) r_{\eta}\left(\phi_{\eta}^{P}\left(x_{3}, b_{2}\right)+\phi_{\eta}^{T}\left(x_{3}, b_{2}\right)\right) \phi_{\omega}\left(x_{2}, b_{2}\right)\right] \\
&\left.\times \alpha_{s}\left(t_{f}^{3}\right) h_{f}^{3}\left(x_{1}, x_{2}, x_{3}, b_{1}, b_{2}\right) \exp \left[-S_{e f}\left(t_{f}^{3}\right)\right]\right\},
\end{aligned}
\end{equation}

\begin{equation}
\begin{aligned}
M_{e}^{P 1}=&-\frac{32}{\sqrt{3}} G_{F} \pi C_{F} r_{\omega} m_{B}^{4} \int_{0}^{1} d x_{1} d x_{2} d x_{3} \int_{0}^{\infty} b_{1} d b_{1} b_{2} d b_{2} \phi_{B}\left(x_{1}, b_{1}\right)\left\{\left[x_{2} \phi_{\eta}^{A}\left(x_{3}, b_{1}\right)\left(\phi_{\omega}^{s}\left(x_{2}, b_{2}\right)-\phi_{\omega}^{t}\left(x_{2}, b_{2}\right)\right)\right.\right.\\
&+r_{\eta}\left(\left(x_{2}+x_{3}\right)\left(\phi_{\eta}^{p}\left(x_{3}, b_{1}\right) \phi_{\omega}^{s}\left(x_{2}, b_{2}\right)+\phi_{\eta}^{T}\left(x_{3}, b_{1}\right) \phi_{\omega}^{t}\left(x_{2}, b_{2}\right)\right)+\left(x_{3}-x_{2}\right)\left(\phi_{\eta}^{P}\left(x_{3}, b_{1}\right) \phi_{\omega}^{t}\left(x_{2}, b_{2}\right)\right.\right.\\
&\left.\left.\left.\left.+\phi_{\eta}^{T}\left(x_{3}, b_{1}\right) \phi_{\omega}^{s}\left(x_{2}, b_{2}\right)\right)\right)\right] \alpha_{s}\left(t_{f}\right) h_{f}\left(x_{1}, x_{2}, x_{3}, b_{1}, b_{2}\right) \exp \left[-S_{c d}\left(t_{f}\right)\right]\right\}.
\end{aligned}
\end{equation}

%\newpage

%\newpage

\end{spacing}
\end{document}